\documentclass[longauth]{aa} 

\usepackage{graphicx}
\usepackage[varg]{txfonts}
\usepackage{lscape}
\usepackage{multirow}
\usepackage{natbib,twoopt}
\usepackage[breaklinks=true, linkcolor=green,citecolor=green,filecolor=cyan,urlcolor=magenta]{hyperref}
\usepackage{xcolor}
\usepackage{siunitx}
\definecolor{hyperlink}{rgb}{0,0,93.3}
\bibpunct{(}{)}{;}{a}{}{,} 
\makeatletter
\newcommandtwoopt{\citeads}[3][][]{\href{http://adsabs.harvard.edu/abs/#3}%
	{\def\hyper@linkstart##1##2{}%
		\let\hyper@linkend\@empty\citealp[#1][#2]{#3}}}
\newcommandtwoopt{\citepads}[3][][]{\href{http://adsabs.harvard.edu/abs/#3}%
	{\def\hyper@linkstart##1##2{}%
		\let\hyper@linkend\@empty\citep[#1][#2]{#3}}}
\newcommandtwoopt{\citetads}[3][][]{\href{http://adsabs.harvard.edu/abs/#3}%
	{\def\hyper@linkstart##1##2{}%
		\let\hyper@linkend\@empty\citet[#1][#2]{#3}}}
\newcommandtwoopt{\citeyearads}[3][][]{\href{http://adsabs.harvard.edu/abs/#3}%
	{\def\hyper@linkstart##1##2{}%
		\let\hyper@linkend\@empty\citeyear[#1][#2]{#3}}}
\makeatother

\def\cm#1{\ifmmode {\,{\rm cm^{-#1}}}                  
	\else \hbox{$\,${\rm cm$^{\rm -#1}$}}\fi}
\def\raw {\ifmmode\rightarrow\else$\rightarrow$\fi}
\def\ex#1{\ifmmode {\times 10^{#1}}         
	\else \hbox{{$\times 10^{\rm #1}$}}\fi}

\newcommand{\kms}{\mbox{km~s$^{-1}$}}

\newcommand{\clam}{\mbox{$\lambda_{\mathrm{c}}$}}
\newcommand{\dlam}{\mbox{$\Delta\lambda$}}

\newcommand{\ntot}{\mbox{$N_{\mathrm{tot}}$}} 

\newcommand{\npol}{\mbox{$N_{\mathrm{pol}}$}}

\newcommand{\mloss}{\mbox{$\dot{M}_{\star}$}}
\newcommand{\mlosssun}{\mbox{$\dot{M}_{\odot}$}}

\newcommand{\ls}{\mbox{$L_{\odot}$}}
\newcommand{\lbol}{\mbox{$L_{\mathrm{bol}}$}}

\newcommand{\msun}{\mbox{$M_{\odot}$}}
\newcommand{\rsun}{\mbox{$R_{\odot}$}}
\newcommand{\mstar}{\mbox{$M_{\star}$}}
\newcommand{\rstar}{\mbox{$R_{\star}$}}
\newcommand{\lstar}{\mbox{$L_{\star}$}}

\newcommand{\rs}{\mbox{$R_{\star}$}}

\newcommand{\teff}{\mbox{$T_{\mathrm{eff}}$}}

\newcommand{\microns}{\mbox{\si{\micro\meter}}}

\newcommand{\gsc}{\object{GSC~07396-00759}}
\newcommand{\bpmg}{$\beta$~Pic~MG}

\newcommand{\degree}{\mbox{$^{\circ}$}}

\begin{document}

	\title{Characterizing the morphology of the debris disk around the low-mass star GSC~07396-00759\thanks{Based on SPHERE observations made with the Very Large Telescope of the European Southern Observatory. Program ID: 1100.C-0481(R).}}
	
	\author{C. Adam\inst{\ref{inst1}, \ref{inst2}, \ref{inst3}}\and 
		J. Olofsson\inst{\ref{inst1}, \ref{inst2}, \ref{inst3}}\and
		R. G. van Holstein\inst{\ref{inst7}, \ref{inst4}}\and
		A. Bayo\inst{\ref{inst1}, \ref{inst2}}\and
		J. Milli\inst{\ref{inst6}}\and
		A. Boccaletti\inst{\ref{inst5}}\and
		Q. Kral\inst{\ref{inst5}}\and
		C. Ginski\inst{\ref{inst18}}\and
		Th. Henning\inst{\ref{inst3}}\and
		M. Montesinos\inst{\ref{inst2}, \ref{inst30}}\and
		N. Pawellek\inst{\ref{inst10}, \ref{inst11}}\and
		A. Zurlo\inst{\ref{inst12},\ref{inst22}, \ref{inst24}}\and
		M. Langlois\inst{\ref{inst22}, \ref{inst23}}\and
		A. Delboulb\'e\inst{\ref{inst6}}\and
		A. Pavlov\inst{\ref{inst6}}\and
		J. Ramos\inst{\ref{inst6}}\and
		L. Weber\inst{\ref{inst9}}\and
		F. Wildi\inst{\ref{inst9}}\and
		F. Rigal\inst{\ref{inst18}}\and
		J.-F. Sauvage\inst{\ref{inst21}, \ref{inst22}}
		}
	
	\institute{
	Instituto de F\'{\i}sica y Astronom\'{\i}a, Facultad de Ciencias, Universidad de Valpara\'{\i}so, Av. Gran Breta\~{n}a 1111, Playa Ancha, Valpara\'{\i}so, Chile
	\\\email{christian.adam84@gmail.com}\label{inst1}
	\and
	N\'{u}cleo Milenio Formaci\'{o}n Planetaria - NPF, Universidad de Valpara\'{i}so, Av. Gran Breta\~{n}a 1111, Valpara\'{i}so, Chile\label{inst2}
	\and
	Departamento de Ciencias B\'{a}sicas, Universidad Vi\~{n}a del Mar, Vi\~{n}a del Mar, Chile\label{inst30}
	\and
	Max Planck Institute for Astronomy, K\"onigstuhl 17, D-69117 Heidelberg, Germany\label{inst3}
	\and 
	Leiden Observatory, Leiden University, PO Box 9513, 2300 RA, Leiden, The Netherlands \label{inst7}
	\and 
	European Southern Observatory, Alonso de Córdova 3107, Casilla, 19001, Vitacura, Santiago, Chile \label{inst4}
	\and
	LESIA, Observatoire de Paris, Universit{\'e} PSL, CNRS, Sorbonne Universit{\'e}, Univ. Paris Diderot,\\ Sorbonne Paris Cit{\'e}, 5 place Jules Janssen, 92195 Meudon, France \label{inst5}
	\and 
	Univ. Grenoble Alpes, CNRS, IPAG, F-38000 Grenoble, France\label{inst6}
	\and 
	Geneva Observatory, University of Geneva, Chemin des Mailettes 51, 1290 Versoix, Switzerland\label{inst9}
	\and 
	Konkoly Observatory, Research Centre for Astronomy and Earth Sciences, Konkoly-Thege Miklós út 15-17, H-1121 Budapest, Hungary\label{inst10}
	\and 
	Institute of Astronomy, University of Cambridge, Madingley Road, Cambridge CB3 0HA, UK\label{inst11}
	\and 
	Anton Pannekoek Institute for Astronomy, Science Park 904, NL-1098 XH Amsterdam, The Netherlands\label{inst18}
	\and 
	DOTA, ONERA, Université Paris Saclay, F-91123, Palaiseau France\label{inst21}
	\and
	N{\'u}cleo de Astronom{\'i}a, Facultad de Ingenier{\'i}a, Universidad Diego Portales, Avenida Ejercito 441, Santiago, Chile\label{inst12}
	\and
	Aix Marseille Universit\'e, CNRS, LAM - Laboratoire d'Astrophysique de Marseille, UMR 7326, 13388, Marseille, France\label{inst22}
	\and
	CRAL, UMR 5574, CNRS, Universit\'e de Lyon, Ecole Normale Sup\'erieure de Lyon, 46 All\'ee d'Italie, F-69364 Lyon Cedex 07, France\label{inst23}
	\and
	Escuela de Ingenier\'ia Industrial, Facultad de Ingenier\'ia y Ciencias, Universidad Diego Portales, Av. Ejercito 441, Santiago, Chile\label{inst24}
	}
	
	\date{Received Month DD, YYYY; accepted July 05, 2021}
	
	
	\abstract
	{
		Debris disks have commonly been studied around intermediate-mass stars. Their intense radiation fields are believed to efficiently remove the small dust grains that are constantly replenished by collisions. For lower-mass central objects, in particular M-stars, the dust removal mechanism needs to be further investigated given the much weaker radiation field produced by these objects.
	}
	{
		We present new observations of the nearly edge-on disk around the pre-main sequence M-type star \gsc, taken with VLT/SPHERE IRDIS in Dual-beam Polarimetric Imaging (DPI) mode, with the aim to better understand the morphology of the disk, its dust properties, and the star-disk interaction via the stellar mass-loss rate.
	}
	{
		We model the polarimetric observations to characterize the location and properties of the dust grains using the Henyey-Greenstein approximation of the polarized phase function. We use the estimated phase function to evaluate the strength of the stellar winds.
	}
	{	
		We find that the polarized light observations are best described by an extended and highly inclined disk ($i\approx 84.3\,\degree\pm0.3$) with a dust distribution centered at a radius $r_{0}\approx107\pm2$\,au.
		Our modeling suggests an anisotropic scattering factor $g\approx0.6$ to best reproduce the polarized phase function $S_{12}$.
		We also find that the phase function is reasonably reproduced by small micron-sized dust grains with sizes $s>0.3$\,\microns. 
		We discuss some of the caveats of the approach, mainly that our model probably does not fully recover the semi-major axis of the disk and that we cannot readily determine all dust properties due to a degeneracy between the grain size and the porosity.
	}
	{	
		Even though the radius of the disk may be over-estimated, our best fit model not only reproduces well the observations but is also consistent with previous published data obtained in total intensity.
		Similarly to previous studies of debris disks, we suggest that using a given scattering theory might not be sufficient to fully explain key aspects such as the shape of the phase function, or the dust grain size.
		Taking into consideration the aforementioned caveats, we find that the average mass-loss rate of \gsc\ can be up to 500 times stronger than that of the Sun, supporting the idea that stellar winds from low-mass stars can evacuate small dust grains in an efficient way.}
	
	\authorrunning{C.~Adam} 
	\maketitle
	%

\section{Introduction}\label{sec:intro}

Debris disks are the natural by-product of the planet formation process. They are second generation, dusty circumstellar disks created by collisions of planetesimals that have formed in previously existing planet-forming disks \citep{2018haex.bookE.165K,2018ARA&A..56..541H}. 
This process releases a large amount of micron-sized dust grains, whose presence either can be inferred through the infrared excess over the photospheric level of their host star, or from spatially resolved scattered (linearly polarized) light observations in the optical or near-infrared.
The average lifetime of those small dust grains, however is much shorter than the typical age of the star, and thus they have to be replenished continuously from the larger bodies. The most efficient process to remove these micron-sized dust grains from the system is usually the pressure exerted by the radiation field of the central star \citep[e.g.,][]{2010RAA....10..383K}. For low-mass stars, however, the irradiation field generally is much weaker, while stellar winds can be strong \citep[see e.g.,][]{2011A&A...527A..57R}. Thus stellar winds are most likely the dominant, yet poorly constrained, mechanism to drive the rapid removal of particles. The question remains poorly addressed because debris disks around low mass stars are rarely detected.

Debris disks are found around about 20\% of A-type stars \citep{2006ApJ...653..675S,2013A&A...555A..11E,2014ApJS..211...25C,2014prpl.conf..521M}, mostly detected through the excess emission in the far-infrared. 
Similar to the searches targeted at solar-type and more massive stars, several large surveys have been conducted to search for cold debris disks around M dwarfs in the mid- and far-infrared \citep{2007ApJ...667..527G,2012A&A...548A.105A,2018MNRAS.476.4584K} as well as the sub-millimeter domain \citep{2006A&A...460..733L,2009A&A...506.1455L}. 
These and other surveys yielded numerous debris disk candidates around young M-stars \citep[for details see e.g.][and references therein]{2020MNRAS.499.3932L}.
However, up to now, only 5 disks around young M-stars have been confirmed based on more than one independent observation, such as spatially resolved imaging (either in scattered light and/or the sub-mm to mm domain): 
AU~Mic \citep{2004Sci...303.1990K,2015Natur.526..230B}, TWA~7 \citep{2016ApJ...817L...2C,2018A&A...617A.109O,2019MNRAS.486.5552B,2020AJ....160...24E,2021ApJ...914...95R}, TWA~25 \citep{2016ApJ...817L...2C}, GJ~581 \citep{2012A&A...548A..86L},Fomalhaut~C \cite[LP 876-10,][]{2021MNRAS.504.4497C}, and, GSC~07396--00759 \citep{2018A&A...613L...6S}. 
Given the low number of spatially resolved detections, we still know very little about debris disks around low-mass stars, and yet several of these disks show very interesting and peculiar features, such as possible spiral-like structures and even gas around TWA~7 \citep{2018A&A...617A.109O,2019AJ....157..117M}, or fast-moving arc-like structures around AU~Mic \citep{2018A&A...614A..52B}.
Debris disks around low-mass stars are also very interesting targets, especially with respect to the possible connection between their occurrence and the possible presence of planets \citep{2011A&A...530A..62R}. 
Although only about 2\% of the M stars have been found to host giant planets \citep{2007ApJ...670..833J}, rocky planets appear to be more frequently detected around low-mass stars, as pointed out by \citet{2015ApJ...807...45D}, who estimated a cumulative planet occurrence rate of $2.5\pm 0.2$ planets per M dwarf with radii $1-4\,R_{\oplus}$ and periods shorter than 200 days. 

However, the reason for the paucity of debris disk detections around low-mass M-type stars remains unclear \citep[see e.g.][]{2020MNRAS.499.3932L}, since stars of all spectral types appear to have a similar detection frequency as protoplanetary disks in the earlier stages of their evolution \citep[see e.g.][]{2005ApJ...631.1134A}. 
In fact, studies of the Lamba Orionis star-forming region carried out on a wide set of spectroscopically confirmed members of the central star cluster Collinder 69 (${\sim}$5--12\,Myr) by \cite{2012A&A...547A..80B}, or the ALMA 887\,\microns\ survey of the disk population in the nearby 2 Myr-old Chamaeleon I star-forming region, conducted by \cite{2016ApJ...831..125P}, indicate that disks around young, low-mass stars with $\mstar\lesssim0.6$\,\msun\ are in fact more frequent than those around higher-mass hosts.
Therefore it is more likely that current observations may simply not be sensitive enough because dust experiences significantly less heating around low-luminosity M dwarfs than around more massive stars. 
Consequently, these colder disks are significantly more difficult to detect at the typical wavelengths (e.g. 24 -- 160\,\microns) that were used to build statistics of debris disks, making the excess emission hard to detect in the spectral energy distribution (SED) of these stars \citep{2014A&A...565A..58M}.
This might increase the chances to miss potential targets when compiling a sample for an observation/survey, and better suited alternatives are needed.

One alternative, as mentioned, are high angular resolution imaging observations in total intensity or polarized light, at optical or near-infrared wavelengths. 
Pioneered by \textit{Hubble Space Telescope} observations \cite[e.g.,][]{2007ApJ...654..595G,2014ApJ...789...58S}, the availability and the recent advances in high contrast imaging instruments, such as the Gemini Planet Imager \cite[GPI,][]{2006SPIE.6272E..0LM} or the Very Large Telescope/Spectro-Polarimetric High-contrast Exoplanet REsearch \citep[VLT/SPHERE;][]{2019A&A...631A.155B} provide direct access to the absorption and scattering properties of the grains and have opened new avenues to detect, resolve, and investigate debris disks at high angular resolution.
Studying how stellar light is scattered off of the dust grains, either using the color of the disk between different bands \citep[e.g.,][]{2008ApJ...684L..41D,2015ApJ...798...96R}, or through studying the phase function over a wide range of scattering angles, i.e. the angle between the star, the dust grain and the observer \citep[e.g.,][]{2016A&A...591A.108O,2017A&A...599A.108M,2019A&A...626A..54M,2019ApJ...882...64R} allows better constraints to be put on the properties of the dust such as their typical grain sizes as well as their porosity, shape, and composition.

The target of our study is \gsc, a young, nearby \citep[71.43$\pm$0.26\,pc,][]{2018A&A...616A...1G} M1-type star.
The star, classified as weak-line T-Tauri star \citep{2011ApJ...740L..17K}, is probably a member of the $\beta$ Pictoris Moving Group \citep[\bpmg, $\approx18\pm2$\,Myr, ][]{2020A&A...642A.179M}, which is known to harbor numerous debris disk host stars like AU~Mic, or CP-72 2713 \citep{2020AJ....159..288M}. 
Total intensity observations of \gsc\ obtained by \cite{2018A&A...613L...6S} have revealed an extended ($r_0=70$\,au) and nearly edge-on disk ($i=83$\degree), probably containing sub-micron-sized grains, a possible indicator of a strong interaction of the stellar radiation field with the disk. 
By comparing their best-fitting model with the observed SED of \gsc\ up to 22\,\microns\ \citep[\textit{WISE}/W4, <4.49\,mJy][]{2014yCat.2328....0C}, the longest wavelength available at this point, \citet{2018A&A...613L...6S} estimated an upper limit for the dust mass in the disk around \gsc\ of $M_{\mathrm{dust}} {\sim} \SI{0.33}{M_{\oplus}}$, corresponding to an upper limit for the fractional luminosity $L_{\mathrm{disk}} / L_\star \leqslant 4 \times 10^{-3}$ due to the lack of far-IR photometry data points.
The disk furthermore displays a small brightness asymmetry, swept-back wings (warps) as well as ripples in the spine of the disk on both sides of the disk.
The available proper motion and radial velocity data suggest that \gsc\ is also likely associated with V4046~Sgr AB, forming a loosely bound hierarchical multiple system  \citep[$a_{\mathrm{proj.}}\approx0.06$\,pc,][]{2006A&A...460..695T,2011ApJ...740L..17K,2018A&A...613L...6S}. V4046~Sgr AB itself a close binary with accretion signatures \citep{2004A&A...421.1159S} and a gas-rich circumbinary disk \citep{2013ApJ...775..136R,2015ApJ...803L..10R}.

In this paper, we aim to compare newly obtained near-infrared polarimetric observations at high angular resolution of \gsc\ with radiative transfer modeling to study the morphology of the disk, and to probe dust properties under the influence of radiation pressure and stellar winds, very rare constraints for low mass stars, and determined to date only for a few debris disks, such as AU Mic \citep{2006A&A...455..987A}, or $\varepsilon$~Eri \citep{2011A&A...527A..57R}.
In Sect.~\ref{sec:obs}, we describe the observations obtained with the SPHERE/ZIMPOL instrument at near-infrared wavelengths.
The model used to analyze the observations is presented in Sect.~\ref{sec:analysis}, followed by
the results reported in Sect.~\ref{sec:results} and their discussion in Sec.~\ref{sec:discussion}. We conclude with a short summary and our conclusions in Sect.~\ref{sec:concl}.

\section{Near-infrared polarimetric imaging}
\label{sec:obs}

\subsection{Observations and data reduction}\label{subsec:obs_and_red}

\begin{table*}[htb!]
	\caption{Summary of the SPHERE-IRDIS observations for \gsc.} 
	\centering	
	\begin{tabular}{lccccccccc}
		\hline\hline\\[-1em]
		Date & DIT & \ntot & \npol & $t_{\mathrm{exp}}$ & Filter & $\tau_{0}$ & Seeing & Am & Strehl \\ 	
		(UTC) & (s) & & & (s) & & (ms) & (\arcsec) & & ($H$) \\[0.5mm]
		\hline	
		2018 Jun 22 & 64.0 & 48 & 12 & 3072 & $H$ & 5.3 & 0.84 & 1.31 & 0.67 \\ 
		
		\hline 
	\end{tabular} \label{table:1}
	\tablefoot{The average DIMM-seeing was measured at $\lambda=500$\,nm. The $H$-band Strehl ratio was measured from the observed IRDIS flux image for \gsc.}
\end{table*}

The observation of \gsc\ took place on 2018 June 22 (UTC, Programme ID: 1100.C-0481(R), P.I.: J.L. Beuzit) as part of the SPHERE Guaranteed Time Observations.
The data were obtained with the SPHERE InfraRed Dual-band Imager and Spectrograph \cite[IRDIS, pixel scale of 12.25\,mas, ${\sim}11\arcsec$\,$\times$\,11\arcsec\, field of view; FoV,][]{2008SPIE.7014E..3LD} using the field-stabilized, dual-beam polarimetric imaging mode \cite[DPI,][]{2014SPIE.9147E..1RL,2020A&A...633A..63D,2020A&A...633A..64V}, and employing the $H$-band ($BB\_H$) filter with a central wavelength \clam=1625\,nm and a width \dlam=290\,nm. To further increase the contrast the Apodized Lyot Coronagraph \citep[mask diameter: 185\,mas,][]{2011ExA....30...39C,2011ExA....30...59G} was used to mask the central star \citep[Hmag=8.76, Rmag=12.01,][]{2003yCat.2246....0C,2017yCat.1340....0Z}. This allows for an efficient suppression of the stellar light which is assumed to be only marginally linearly polarized, while keeping the scattered, i.e. polarized, light from the circumstellar disk relatively unaffected. 

Each polarimetric observation consists of a set of four linear-polarization components, called Stokes $Q^{+}$, $Q^{-}$, $U^{+}$, and $U^{-}$, that are obtained by subtracting the two beams with orthogonal polarization states recorded simultaneously on the detector and tuning their polarization direction with a half-wave plate (HWP) with positions of 0\degr, 45\degr, 22.5\degr, and 67.5\degr, respectively. 
We took one exposure for each of the Stokes $Q^{+}$, $Q^{-}$, $U^{+}$, and $U^{-}$ components, each with a detector integration time (DIT) of 64\,s. The polarization cycle of $Q^{+}$, $Q^{-}$, $U^{+}$, and $U^{-}$ was then repeated twelve (\npol) times, adding up to a total of 48 exposures with a total integration time ($t_{\mathrm{exp}}$) of 51.2 minutes. An overview of our observations is presented in Table~\ref{table:1}. 
We also list the observing conditions at the time of observation, such as the average coherence time ($\tau_{0}$), the average seeing, estimated from the Differential Image Motion Monitor (DIMM), the airmass (Am) and the Strehl ratio in the $H$-band, measured from the flux image.

Since the star is obscured by the coronagraphic mask in the science images, a flux and center calibration frame were taken in addition to the science observations. The flux calibration frames were taken with the central star moved away from the coronagraphic mask. 
The center calibration frames were taken after the star was aligned behind the coronagraphic mask and with the deformable mirror system \cite[SAXO,][]{2006OExpr..14.7515F} used to introduce a waffle pattern to create equidistant calibration spots outside of the coronagraphic mask. 
This allows us to accurately determine the stellar position behind the coronagraph from the center image while we compute the Strehl ratio in the $H$-band for \gsc\ directly from the flux image with the star off-centered from the coronagraph.
The $H$-band Strehl ratios estimated by the adaptive optics system during the observation are recorded in separate FITS files\footnote{``Classified as OBJECT, AO'' in the ESO data archive} (called GEN-SPARTA data).
From these data we find Strehl ratios for \gsc\ ranging from 0.6 to 0.77, and in good agreement with our measured Strehl ratio of 0.67 from the flux image.

The data were reduced using the IRDAP\footnote{\url{https://irdap.readthedocs.io}} \citep[IRDIS Data reduction for Accurate Polarimetry,][]{2020A&A...633A..64V,2020A&A...633A..63D} pipeline. 
IRDAP is a dedicated pipeline for the reduction of polarimetric data obtained with IRDIS, capable of differentiating and correcting instrumental and stellar polarization. 
For a detailed description of the reduction procedure and a discussion of the applied corrections we refer the reader to \cite{2020A&A...633A..64V} and \cite{2020A&A...633A..63D}. The reduction can be summarized as follows.

After applying the standard calibration routines, including sky-frame subtraction, flat-fielding and bad-pixel correction, the images are split into two individual frames representing the left and right sides of the IRDIS detector, corresponding to the parallel and perpendicular polarized beams, respectively. Then, the precise position of the central star is measured using the star center calibration frames on both image sides separately, and the right side of the image is shifted to a common center and subtracted from the left side.
The pipeline then applies the double-difference method \citep[see, e.g.][]{1996aspo.book.....T} to obtain the linear Stokes parameters $Q$ and $U$ corrected for instrumental polarization created downstream of the HWP and the corresponding total-intensity images computed with the double sum, respectively. 
However, the double difference does not remove instrumental polarization caused by the telescope and instrument mirrors upstream from the HWP which is assumed to be proportional to the total intensity image, as shown in \cite{2011A&A...531A.102C}, nor does it remove the most important cross talk contributions \citep{2020A&A...633A..63D}. 
IRDAP uses a Mueller matrix model to determine the polarimetric response function for the polarimetric imager \citep{2020A&A...633A..63D} and to correct for these instrumental polarization effects.
This model describes the complete optical path of SPHERE/IRDIS, i.e., telescope and instrument, and has been fully validated with measurements using SPHERE’s internal source and observations of unpolarized standard stars \citep{2020A&A...633A..64V}.
The images of Stokes $Q$ and $U$ incident on the telescope are computed by setting up a system of equations describing every measurement of $Q$ and $U$ and solving it -- for every pixel individually-- using linear least-squares. The $Q$ and $U$ images thus created, however, may still contain some stellar polarization that is constrained by measuring the flux in the $Q$ and $U$ images around regions that should be virtually devoid of polarized signal from the disk. 
The final product of the reduction pipeline are the images of the azimuthal Stokes parameter $Q_{\phi}$ and $U_{\phi}$ \citep[for definition see][]{2020A&A...633A..63D}, where $Q_{\phi} > 0$ is equivalent to a azimuthal polarization component (with respect to the position of the star), $Q_{\phi} < 0$ to a radial component, and $\pm U_{\phi}$ signal is equivalent to polarization angles oriented at $\pm 45$\degree with respect to the azimuthal component, respectively.

\subsection{Observational results}\label{subsec:obs_res}

\begin{figure*}[htbp!] 
	\includegraphics[width=180mm]{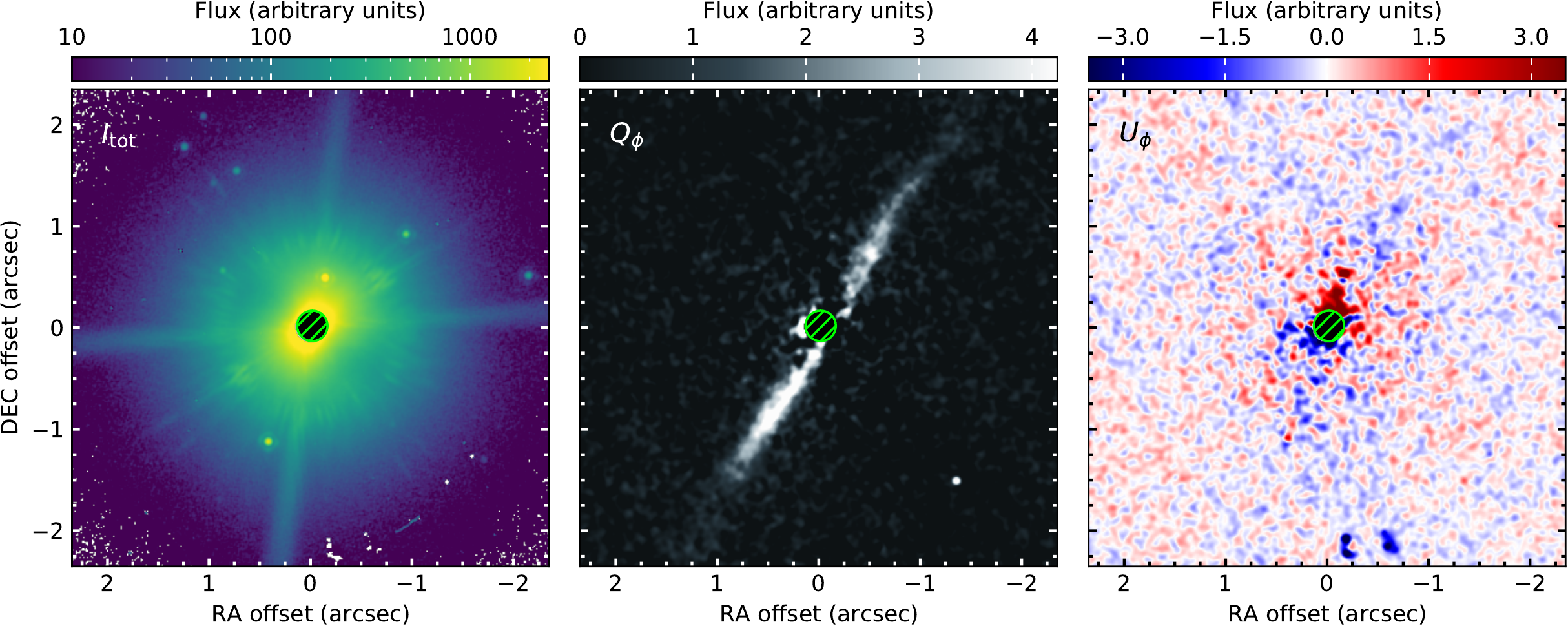}
	\caption{SPHERE/IRDIS polarimetric imaging observations of \gsc. Shown from left to right are the total intensity $I_{\mathrm{tot}}$ (log stretch) as well as the $Q_{\phi}$ and $U_{\phi}$ images (both in linear stretch). The $Q_{\phi}$ and $U_{\phi}$ images were convolved with a Gaussian PSF ($\sigma=2\,$pixel) to increase the visibility in this plot. The coronagraphic mask is indicated by the green circular region in each panel. North is to the top and east to the left in each panel. Please note that the disk is not visible in the total intensity image (left panel). The structure appearing to extend from the NW to the SE is a PSF artifact.}
	\label{fig:data_obs}
\end{figure*}

\begin{figure}[htbp!] 
	\centering
	\includegraphics[width=80mm]{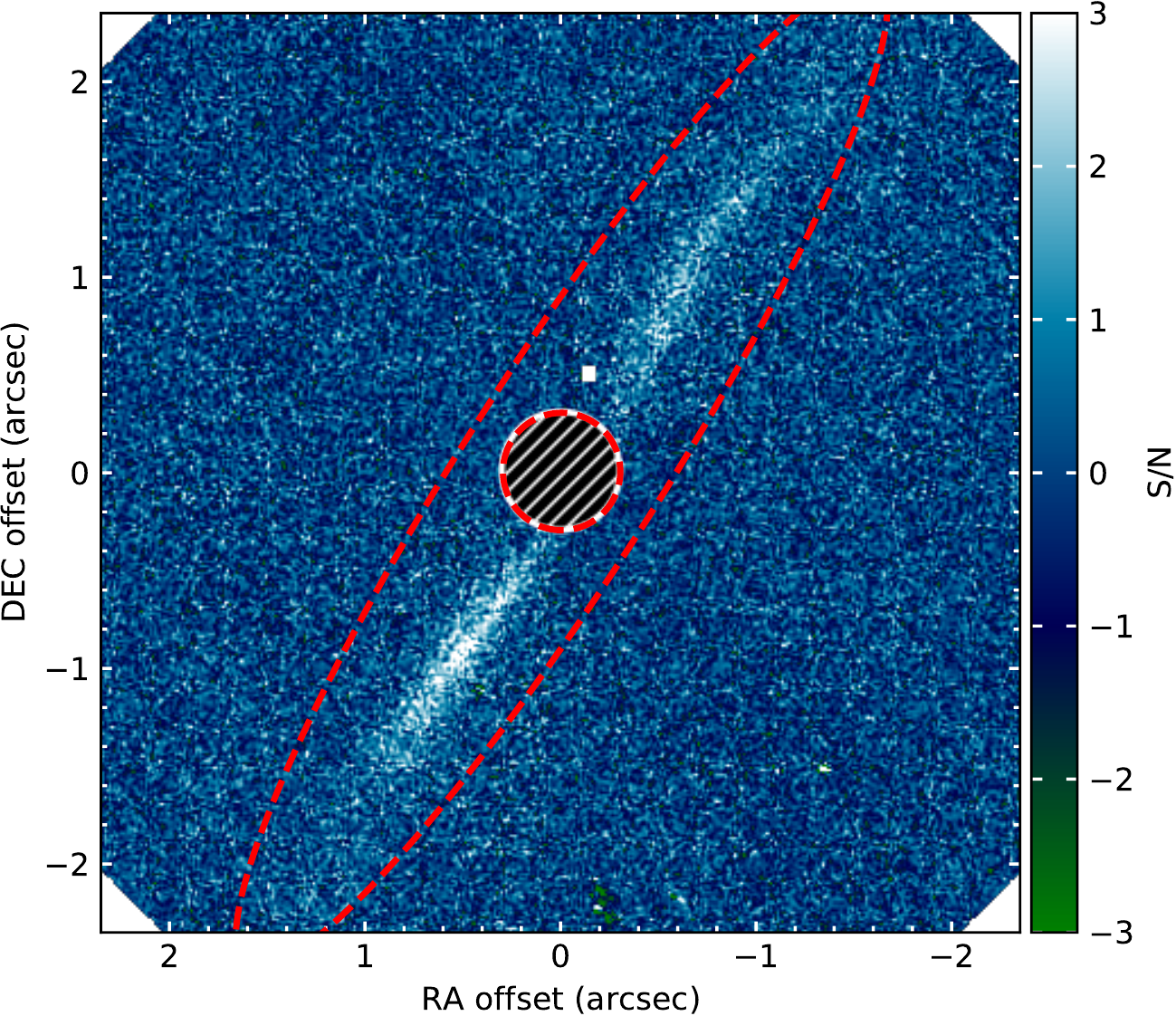}
	\caption{Signal-to-noise map estimated from the $Q_{\phi}$ and noise images (without convolution) of our IRDIS DPI observations. The plot is shown with a linear stretch between [-3$\sigma$, 3$\sigma$]. The inner and outer model boundaries are indicated by the circle and the ellipse in the image. North is to the top and east to the left.}
	\label{fig:data_snr}
\end{figure}

Figure~\ref{fig:data_obs} shows the final collapsed total-intensity, $Q_{\phi}$ and $U_{\phi}$ images of our IRDIS DPI $H$-band observations of \gsc. 
As can be seen, the total-intensity image (left panel) shows numerous point sources, and in the FoV (cropped to ${\sim}\,2.2\arcsec \times 2.2\arcsec$) of our IRDIS observation alone, we identified more than 10 different sources. 
In fact, \cite{2018A&A...613L...6S} detected a total of 109 point sources in their dual-band imaging observations taken with IRDIS as part of the SHINE survey \citep{2020arXiv200706573V}, in the night of June 15, 2017. 
However, by using observations of \gsc\ taken in the International Deep Planet Survey \citep[IDPS;][]{2012A&A...544A...9V,2016A&A...594A..63G} with the NIRC2 instrument at Keck, and with a time difference of about 10 years, they were able to identify 70 objects as background objects based on common proper motion. Another 32 sources could be rejected as companions based on their location in the color-magnitude diagram (CMD).
For the 7 remaining objects the results could not be determined unambiguously, however all but one have separations above 5\arcsec\ (a$_{\mathrm{proj.}}>400$\,AU), and the colors of those objects are similar to that of the identified background objects \citep{2018A&A...613L...6S}.
It is thus highly likely that these are unrelated background objects too. 

The middle-panel of Fig.~\ref{fig:data_obs} shows the reduced $Q_{\phi}$ image. 
Noticeable is the almost edge-on disk extending about 1.3\arcsec\ to the South-East (SE) and North-West (NW). 
Similar to the findings by \cite{2018A&A...613L...6S}, we detect an asymmetric brightness distribution, with the SE side of the disk appearing slightly brighter than the NW side. The disk also appears to be warped on the NW side, indicating a more complicated structure of the disk. 
The image however shows one of the major advantages of polarimetric observations. 
That is, since stellar light is usually not polarized, the image is not contaminated by background stars which makes it easier to detect circumstellar matter. In fact, except for one source/artifact located about 0.5\arcsec\ North of the central star, the disk appears to be free of contamination from other sources visible in the total intensity image.
We masked this source separately during our subsequent analysis to minimize its contribution. 

We also note that on average the $U_{\phi}$ image (Fig.~\ref{fig:data_obs}, right panel) is mostly devoid of signal. This is consistent with what we expect in case of single scattering in a centrally illuminated, optically thin disk.
We, thus, assume that the $U_{\phi}$ image to the first order contains only noise and measure these uncertainties from the standard deviation in concentric annuli with a width of 1 pixel in the $U_{\phi}$ image to create a noise map. This allows us to estimate the goodness of fit for our models as well as to create the signal-to-noise ratio (S/N) map shown in Fig.~\ref{fig:data_snr}, computed as the ratio of $Q_{\phi}$ image and the estimated noise map. To minimize the influence of the higher noise close to the central star on the modeling, we introduce a circular numerical mask with $r=0.3$\arcsec\ as inner boundary as well as an ellipse as outer limit for the modeling, which are also indicated in Fig.~\ref{fig:data_snr}. Values inside the circular mask and outside the ellipse are not considered in the modeling.

\subsection{Stellar properties}\label{subsec:stellar_prop}

\begin{table}[hbtp!]
	\caption{Summary of stellar parameters of \gsc.} 
	\label{table:2}	
	\centering	
	\begin{tabular}{l c c}
		\hline\hline\\[-1em]
		Parameter                 & Value          & Reference \\
		\hline 
		SpT                       & M1Ve           & a \\[0.5mm]
		\teff\,[K]                & 3800$\pm$100    & - \\
		$D$\,[pc]                 & 71.43$\pm$0.26 & b \\
		\lstar\,[\ls]             & ${\sim}$0.13  & - \\
		\rs\,[\rsun]              & ${\sim}$0.71  & - \\
		RV\,[\kms]                & -5.7$\pm$0.8   & c \\
		$v \sin i $ [\kms]        &  3.0$\pm$1.5 & c \\
		$P_{\mathrm{rot}}$\,[d]   & 11.63$\pm$0.02 & - \\
		Age [Myr] & $\lesssim$20 & - \\
		$M_{\star}$\,[$M_{\sun}$] & 0.62$^{+0.04}_{-0.02}$   & - \\[0.5mm]
		\hline 
	\end{tabular}
	\tablefoot{SpT: Spectral type. $D$: Distance to the Sun. \teff: Effective temperature. \lbol: Bolometric luminosity. \rs: Stellar radius. $M_{\star}$: Stellar mass. RV: Radial velocity. $P_{\mathrm{rot}}$: Rotation period. (a): \citet{2013ApJS..208....9P}; (b): \citet[Gaia DR2,][]{2018A&A...616A...1G}; (c): \citet{2014ApJ...788...81M}; (-): This work.}
\end{table}

Along with the observations of \gsc, and a thorough analysis for possible companions, \cite{2018A&A...613L...6S} collected and estimated various stellar parameters, and we refer the reader to their paper for a detailed description of these parameters.
Nevertheless, we revised the most important parameter, i.e. the distance to \gsc, using new astrometric data from the Gaia satellite mission data release 2 \citep[Gaia DR2,][]{2018A&A...616A...1G}. In Table~\ref{table:2}, we summarize the resulting stellar parameters which we estimated as follows.

Employing the VO SED Analyzer tool\footnote{Version 7 (soon to be released), \url{http://svo2.cab.inta-csic.es/theory/vosa/}}
\citep[VOSA,][]{2008A&A...492..277B}, we first collected photometric data ranging from the ultraviolet (UV) at ${\sim}0.297$\,\microns\ \citep[\textit{XMM-OM}/UVW1, 0.38\,mJy,][]{2012MNRAS.426..903P} to the mid-infrared (MIR) wavelengths, with the longest wavelength being 22\,\microns\ at this point. We then applied a BT-Settl-CIFIST model grid of theoretical spectra \citep{2015A&A...577A..42B} to derive the stellar parameters from the constructed SED.
The data indicate a small UV excess which we account for by reducing the weight of the UV contribution to the fit. Furthermore allowing the interstellar extinction ($A_{\mathrm{v}}$) to vary between 0 and 0.4, we find an effective temperature of $\teff=3800\pm100$\,K and a stellar luminosity of $\lstar=0.132$\,\ls\ for \gsc. From the position in the Hertzsprung-Russell diagram (HRD) combined with theoretical isochrones and mass tracks \citep{2015A&A...577A..42B} we estimate an upper limit for the age of $\lesssim$\,20\,Myr and a stellar mass of ${\sim}0.62\,{\msun}$ (but with a flat distribution between 0.6 and 0.68\,\msun). 
From the dilution factor of the SED fit, combined with the distance of 71.43\,pc, we estimate a stellar radius of $\rstar\sim0.71\,\rsun$.

When compared with the previous estimates by \citet{2018A&A...613L...6S}, we find a slightly higher effective temperature, but a significantly smaller stellar radius for \gsc.
This, however, is not entirely surprising if we take into account that i) the luminosity, and thus the stellar radius, depends on the distance to the star, and ii) in particular young stars that are members of young moving groups, such as the \bpmg, appear over-luminous, and hence inflated, when compared to older (field) stars \citep{2014ApJ...792...37M}.
This, combined with the choice of the model grid could explain the discrepancy in the estimated stellar properties between our results and those obtained by \citet{2018A&A...613L...6S}.

Additionally, we analyze available light-curves of \gsc\ observed with the Transiting Exoplanet Survey Satellite \citep[TESS,][]{2015JATIS...1a4003R}, and during the All-Sky Automated Survey for Supernovae \citep[ASAS-SN,][]{2014ApJ...788...48S,2017PASP..129j4502K}, respectively, to measure the stellar rotation period. \gsc\ was observed with TESS (600--1000\,nm), covering an observation period of about 27 days, between 2019-06-19 and 2019-07-17, whereas the ASAS-SN light curve was obtained in the $V$-band over a time period of 2.5 years between 2016-03-10 and 2018-09-22. Using Lightkurve, a Python package for Kepler and TESS data analysis \citep{2018ascl.soft12013L}, we estimate a stellar rotation period $P_{\mathrm{rot}}=11.63\pm0.02$\,d from the Lomb–Scargle periodogram \citep{1976Ap&SS..39..447L,1982ApJ...263..835S} of the TESS light curve (see Fig.~\ref{fig:lc_tess}), and $P_{\mathrm{rot}}=12.06\pm0.02$\,d from the ASAS-SN light curve (see Fig.~\ref{fig:lc_asas-sn}), respectively. The uncertainty associated with each period was estimated by re-sampling the light curve using the bootstrap method (with replacement), and corresponds to the 95\% confidence interval of our sample estimates. We note, however, that these are only statistical uncertainties and they do not represent the totality of the error budget of the light curves. For example, the errors do not reflect properly that the TESS light curve only covers about two periods, or that the period estimates are probably affected by two additional sources ($\Delta\mathrm{mag}=2$) that fall on the same TESS pixel as \gsc, given the pixel scale of 21$\arcsec$ per pixel.
Nevertheless, both period estimates are in good agreement with the period of $P_{\mathrm{rot}}=12.05\pm0.5$\,d reported by \citet{2017A&A...600A..83M}. 
From the period estimated from the TESS light curve, together with the evaluated radius ($\rstar{\sim}0.71\,\rsun$) and a $v \sin i=3.0\pm1.5$\,\kms \citep{2014ApJ...788...81M} we estimate a stellar inclination of $i{\sim}80\degree\pm20$\degree\ \citep[see e.g. Eq.~1;][]{2020A&A...642A.212J}, consistent with a stellar rotation that is likely co-planar with that of the disk, as proposed by \citet{2018A&A...613L...6S}.
Very few disks have a measured stellar inclination to which the disk inclination can be compared \cite[see e.g.][]{2014MNRAS.438L..31G}. Our results therefore might be helpful in future studies of the relation between the stellar inclination and the inclination of the disk.

As mentioned above, the model fit of the SED of \gsc\ suggests a small UV excess. We also found that \gsc\ has various detections in the X-ray band, reported in the fourth generation of serendipitous source catalogs \citep[4XMM,][]{2020A&A...641A.137T}. 
The detected release of energy in the X-ray and UV portion of the stellar spectrum suggests that \gsc\ is subject to strong coronal emission, and possibly intense and frequent flares, common for young and low-mass stars such as \gsc. 
These stellar flares are caused by the re-connection of magnetic field loops on the surface of the star \citep{2019ApJ...870...10M}.
How frequently these stellar flares occur also depends on how magnetically active the star is.
For example, AU Mic which is similar in luminosity and age to \gsc\ (both are members of the \bpmg) is known to frequently present X-ray and EUV flares at a rate of about 0.9 flares per hour, during which the EUV and X-ray luminosities increase by a typical factor 10 \cite[see, e.g.,][and references therein]{2006A&A...455..987A}.
In fact, recent studies of the flare-activity of young ($\leq100$\,Myr) K- and M-stars in the Upper Sco region using Kepler K2 data found that early and late-type M-stars might have 10000 to 80000 times as many high-energy flares, so-called super-flares with $E\geq5\times10^{34}$\,erg, than solar like stars \citep{2019arXiv191109922G}.
Such an enormous activity will certainly affect the direct environment around the stars, especially potential disks and/or planets that orbit the star.

Although we do not detect flares in the available light curves of \gsc, such an increased flare-rate might also be the case for \gsc, because not only is the star presumably young ($\approx20$\,Myr), but it is likely also magnetically active, as indicated by the detection of the H$\alpha$ and H$\beta$ lines in emission \citep{2018A&A...613L...6S}, and the moderate rotation period of $P_{\mathrm{rot}}\approx12$\,d. 
Thus, even if \gsc\ is currently in a quiescent state, the present coronal activity of the star and its effect on the disk may not be negligible, given the observed variations in the X-ray flux and the apparent UV excess detected in \gsc.

\section{Analysis} \label{sec:analysis}

Besides gravity, the orbital parameters of small \microns-sized particles, observed in the visible and near-infrared using scattered and linear-polarized light, are mainly affected by radiation pressure and stellar winds. In particular the latter can dominate the effects on circumstellar grains in debris disks around late-type, low-mass stars as first pointed out by \citet{2005ApJ...631.1161P}.
Hence, in this study we try to constrain the dust properties, as well as the distribution of dust grains in the debris disk around \gsc, taking into account not only radiation pressure but also the effect from the stellar wind of the central star.

\subsection{Code description} \label{subsec:rtm}

The code used in this study is based on the same code as was described in \cite{2019A&A...630A.142O}.
However, since we have made some changes to the code, in the following, we give a short summary of the code description and the modifications we made to also account for the effects of stellar wind pressure on the dust grains. 
	
The observed dust is released from a belt of planetesimal-sized parent bodies (their numbers, sizes and masses do not need to be specified) which is defined by 6 parameters: a reference radius $r_{0}$ which is the semi-major axis of the orbit, eccentricity $e$, inclination $i$, argument of periapsis $\omega$, the position angle on the sky $\varphi$, measured positive from north to east, and the width of the ring $\delta_{r}$.
All the dust grains originate from those parent bodies that a priori, all share the same $e$ and $\omega$, and have a radial distribution that follows a normal distribution centered at a radius $r_{0}$ with a standard deviation of $\delta_{r}$.
The dust grain-size distribution, used in the code, follows a differential power-law 
\begin{eqnarray}
dn(s) \propto s^{-p} ds \, , \; \label{eq:dn_s}
\end{eqnarray}
where $s$ is the grain size, and $p=3.5$ is the slope index of a particle distribution derived from an idealized collisional cascade, following \citet{1969JGR....74.2531D}.
This distribution is divided in $n_{\mathrm{g}}$ intervals (equidistant in logarithmic space between $s_{\mathrm{min}}$ and $s_{\mathrm{max}}$, respectively), and the number density of grains in each bin is computed using Eq.~(2) of \cite{2008A&A...487..205D}. 

For each grain size the code then computes the dimensionless parameter $\beta$ which, in previous versions of the code, was defined as the ratio between the radiation pressure and the gravitational forces \citep{1979Icar...40....1B}. 
If this ratio exceeds 0.5 for an initially circular orbit, the dust particle gets pushed into a hyperbolic orbit and leaves the system \citep{2006A&A...455..509K}. Hence, for a given grain composition and porosity, this $\beta$ parameter cut-off can be used as a very close approximation of the effective radius below which dust particles will leave the system \citep{2019AJ....157..157A}.
Following the investigation of AU~Mic \citep{2006A&A...455..987A,2017A&A...607A..65S} we modified the code to also include effect of the stellar wind pressure force on the dust grains.
The net pressure force acting on a grain is then defined by $\beta=\beta_\mathrm{RP}+\beta_\mathrm{SW}$ \citep{2017A&A...607A..65S}, where $\beta_{\mathrm{RP}}$ is the ratio between the radiation pressure forces and the gravitational forces, and $\beta_{\mathrm{SW}}$ is the ratio between the wind pressure and gravitational forces.
The individual contributions of the two pressure forces to $\beta$ can be estimated via
\begin{eqnarray}
\beta_{\mathrm{SW}} &=& \frac{3}{32\pi} \frac{\mloss V_{\mathrm{SW}} C_{\mathrm{D}} }{G \mstar \rho s} \, , \label{eq:b_sw}
\end{eqnarray}
where \mloss\ is the stellar mass-loss rate, $V_{\mathrm{SW}}$ is the stellar wind speed, $C_{\mathrm{D}}$ the dimensionless free molecular drag coefficient which we take equal to 2 \citep[see e.g.][and references therein]{2006A&A...455..987A}, $G$ the gravitational constant, \mstar\ the mass of the star, and $\rho$ the volumetric mass density of the dust, respectively, and
\begin{eqnarray}
\beta_{\mathrm{RP}} &=& \frac{3}{16\pi} \frac{\lstar \langle Q_{\mathrm{RP}} \rangle}{c G \mstar \rho s} \, , \mathrm{with } \, \langle Q_{\mathrm{RP}} \rangle = \frac{\int_{\lambda} Q_{\mathrm{RP}} F_{\lambda} d\lambda}{\int_{\lambda} F_{\lambda} d\lambda} \, , \label{eq:b_pr}
\end{eqnarray}
where \lstar\ the stellar luminosity, $Q_{\mathrm{RP}}$ the dimensionless radiation pressure efficiency, which along the wavelength also depends on the dust grain size and composition, $F_{\lambda}$ the stellar flux, and $\lambda$ the wavelength.
As can be seen from the equations, in general and for sufficiently large dust grains $\beta$ varies with $s^{-1}$. It should be noted however that for smaller grain sizes, the relationship between $\beta$ and $s$ becomes more complex and increasingly dependent on the grain composition, the stellar mass-loss rate \mloss, and stellar wind speed $V_{\mathrm{SW}}$ \citep{2017A&A...607A..65S}.

For a given $\beta$ the code draws $3\,000$ samples from an uninformative uniform prior distribution for the mean anomaly, to decide where the dust grain is located upon its release. For each realization, the code then calculates the \textquotedblleft updated\textquotedblright orbital parameter ($a_n$, $e_n$, and $\omega_n$) using Eq.~(2) in \cite{1999ApJ...527..918W,2006ApJ...639.1153W,2016ApJ...827..125L}.
In case the updated eccentricity $e_{\mathrm{n}}$ is larger or equal to zero and strictly smaller than unity (to avoid hyperbolic orbits), the resulting orbits are populated with 300 dust particles, uniformly distributed in mean anomaly. 
The vertical distribution of the disk is accounted for by drawing from a normal distribution with a standard deviation $h=0.04\times r$, following \citet{2009A&A...505.1269T}. 
This allows us to account for column density effects, as explained in further detail in \cite{2020A&A...640A..12O}.
The $(x,y,z)$ positions of each particle are registered, and depending on inclination and position angle of the disk, the corresponding closest pixel is determined, thus producing number density maps for each value of $\beta$.

Furthermore, and as discussed in \cite{2019A&A...630A.142O}, when computing the number density maps for each $\beta$ value, the contribution of each particle is also multiplied by a correction factor \citep{2006ApJ...648..652S,2016ApJ...827..125L}, which is roughly proportional to their total orbital period divided by the time spent within the birth ring.
This correcting “enhancement” factor to the high-$\beta$ grain number density allows us to at least to the first order account for the fact that small grains produced inside the belt on high eccentricity (bound) orbits will spend most of their time in the collision-free outer regions where they cannot be collisionally destroyed, hence enhancing significantly their collisional lifetimes, and thus their number density \citep{2006ApJ...648..652S,2008A&A...481..713T}. 

Once each of the $3000\times 300$ particle has been launched the code computes the scattering angle between the central star and the observer for each particle in the image. 
The polarized light images per grain size bin are then computed by multiplying the estimated number density of each pixel by $S_{12}\times \pi s^{2}\times Q_{\mathrm{sca}}/(4\pi r^2)$, where $r$ is the distance from the particle to the star, $Q_{\mathrm{sca}}$ the scattering efficiency, and $S_{12}$ the polarized phase function.
The final image is the collapse of all individual images for each grain size bin, weighted by the grain size distribution $n(s)$.

\subsection{Modeling strategy}

\subsubsection{Caveats and modeling approach}

Most likely due to a degeneracy between the minimum grain size and the porosity of the dust particles or the high inclination of the disk around \gsc\ our primary modeling attempts were not converging on a unique solution (see Sec.~\ref{sec:results} for a detailed discussion).
To simplify the problem, we therefore used a different, two-step approach in which we first focused on the morphology of the disk and then on the dust properties. The procedure can be summarized as follows.

We first try to alleviate the influence of some of the dust properties by replacing the polarized phase function in our disk model with an analytical form, the Henyey-Greenstein approximation \citep{1941ApJ....93...70H} described in Sec.~\ref{subsec:descr_disk}.
This form of the phase function is parameterized for all grain sizes.
Thus, the absolute value of $\beta$ for each grain size becomes less relevant, and the spatial distribution of the dust particles is mostly determined by the global shape of the $\beta$ function ($\beta(s) \propto s^{-1}$).
By sampling the entire range of $\beta$ values below 0.5, we can therefore estimate the spatial distribution of the dust particles in the disk, without having to pay too much attention to their actual sizes and properties.

{Outside of the birth ring there is also an over-abundance of small dust grains, mostly due to the effect of radiation pressure, as discussed for example in \cite{2008A&A...481..713T}.
While the grain size distribution is expected to follow the "classical" distribution with an exponent $p=3.5$ (see above) in the birth ring, outside of the birth ring, the smallest dust grains have an increased collisional lifetime as they are set on highly eccentric orbits and, thus, spend most of their time near their apoapsis, i.e. they survive longer and hence contribute even more to the flux observed in scattered light. In fact, the deviation of the size distribution from -3.5 can actually be quite significant \citep[see e.g. Fig.~3,][]{2006ApJ...648..652S}, resulting in the small dust grains dominating outside of the birth ring. Therefore the phase function we retrieve is the one for the dominant/most representative grain size, which we assume to be close to the blow out size of the dust particle.

With the radial distribution and the phase function established by our disk model, we can then, in the second step, try to conversely infer other properties of the dust particles. 
We achieve this by comparing the inferred analytical polarized phase function with a model dependent polarized phase function computed by other means such as effective medium theory (EMT), the discrete dipole approximation \citep[DDA;][]{1973ApJ...186..705P,1988ApJ...333..848D}, or a distribution of hollow spheres \citep[DHS, see][]{2005A&A...432..909M}. 
Depending on the grain composition and porosity we can then try to estimate individual parameters, such as the effective radius ($s_{\mathrm{min}}$), below which particles will not stay in the system, or the stellar mass-loss rate (\mloss).

\subsubsection{Modeling the disk}\label{subsec:descr_disk}

The free parameters of our disk model are the reference radius of the radial distribution $r_{0}$, its standard deviation $\delta_{r}$, the inclination $i$, and the position angle $\phi$. 
The high inclination of the disk around \gsc, however, makes it hard to reasonably constrain the eccentricity. We therefore assume a circular disk and fix the eccentricity to $e=0$, making the argument of periapsis $\omega$ irrelevant for the modeling. For the stellar parameters we use a bolometric luminosity of 0.13\,\lstar, a stellar mass of 0.62\,\msun, and a distance of 71.43\,pc.

We sample the grain size distribution so that the entire range of $\beta$ values from 0.5 down to $10^{-5}$ is probed, i.e. the dust particles remain gravitationally bound over the entire grain size distribution between $s_{\mathrm{min}}$ and $s_{\mathrm{max}}$.
To determine the polarized phase function $S_{12}$ as a function of the scattering angle, we use a modification of the analytical Henyey-Greenstein approximation \citep[HG,][]{1941ApJ....93...70H} which is given by
\begin{eqnarray}
S_{\mathrm{12,HG}} &=& \frac{1-g^2}{4\pi(1+g^2-2g\cos(\theta))^{3/2}} \frac{1-\cos^2(\theta)}{1+\cos^2(\theta)} \, , \label{hg}
\end{eqnarray}
where $\theta$ is the scattering angle and $g$ the anisotropic scattering factor ($-1\leq g \leq 1$), which governs the scattering efficiency as a function of the scattering angle $\theta$.
The first factor in Eq.~\ref{hg} is the HG function that describes the scattered flux produced by the photons hitting and interacting with the dust particles, while the second term in Eq.~\ref{hg} takes into account the angle dependence of the linear polarization produced by the particle scattering, using the Rayleigh scattering function as a simple approximation. 
Thus, for $g=0$ (isotropic scattering) the maximum of scattered polarized flux occurs at $\theta=90\,\degree$, while for $g>0$ the maximum is shifted to smaller scattering angles resulting in an asymmetry in the amount of polarized light received from the front and backsides of the disk \citep{2017A&A...607A..90E}.

The best solution to our model with 5 free parameters ($r_{0}$, $\delta_{r}$, $i$, $\phi$, and $g$) is determined using an affine invariant ensemble sample Monte-Carlo Markov Chain \citep[$\mathtt{emcee}$ package,][]{2013PASP..125..306F}. 
For the initial conditions of the model we use the disk parameters ($r_{0}$, $i$, and $\phi$) reported by \cite{2018A&A...613L...6S}. We then draw random samples from uniform priors we also report in Table~\ref{table:3}, using a Monte Carlo Markov chain composed of 30 walkers, a burn-in phase of 1000 models, followed by the actual modeling using chains of 10000 models for each walker. 
To speed up the modeling process, we exclude all image data points located outside of an elliptical mask defined by a semi-major and semi-minor axis of 2.3\,\arcsec and 0.6\,\arcsec, and a position angle of 150\degree, respectively. We also place a numerical mask with a radius of 0.3\,\arcsec on the center to reduce the influence of the increased noise around the central star on the modeling (see Fig.~\ref{fig:data_snr}).
The goodness of fit for each model is estimated as the sum of the squared residuals divided by the uncertainties.
The results of the disk modeling are presented in Sec.~\ref{sec:results}. 
	
\subsubsection{Polarized phase function}\label{subsec:pp_ml}

In the second step of our analysis we use the parameterized polarized phase function obtained in the previous step to try to infer further dust properties by comparing it to a model-dependent phase function.
In this study, we use the $\mathtt{OpacityTool}$, a dedicated Fortran package\footnote{\url{https://dianaproject.wp.st-andrews.ac.uk/data-results-downloads/fortran-package}} of the DIANA project \cite[see][]{2016A&A...586A.103W,1981ApOpt..20.3657T}, to compute dust opacities for the purpose of disc modeling.
The package uses a distribution of hollow spheres \citep[DHS, see][]{2005A&A...432..909M} to compute absorption and scattering properties, as well as six different elements of the scattering matrix, including $S_{11}$, the scattering phase function, and $S_{12}$, the polarized phase function, respectively.
We use the same grain size distribution as in the previous step to calculate polarized phase functions.

Preliminary tests, however, suggested that a size distribution between $s_{\mathrm{min}}$ and $s_{\mathrm{max}}$ ($s_{\mathrm{max}} >> s_{\mathrm{min}}$) could not match the phase function derived from the modeling of the disk, given the limitations of the data and the observed degeneracies in the model. Therefore, we chose to only compare our results to models with one representative grain size to find the typical grain size that can best explain the inferred polarized phase function. For each grain size we calculate polarized phase functions varying the porosities from 0.01 to 0.99 in steps of 0.01, and the grain size ($s$) between $0.01\,\microns$ ($s_{\mathrm{min}}$) and 1\,mm ($s_{\mathrm{max}}$), respectively.
The $\mathtt{OpacityTool}$ calculates the opacities and other properties using a mixture of 85\% amorphous laboratory silicates \citep[][Mg$_{0.7}$Fe$_{0.3}$SiO$_{3}$]{1995A&A...300..503D} with 15\% amorphous carbon \citep[][BE-sample]{1996MNRAS.282.1321Z}.
The effective refractory index of the porous material is calculated by applying the \cite{1935AnP...416..636B} mixing rule. The maximum hollow volume ratio is set to $f_\mathrm{max}=0.8$.
This filling factor $f_\mathrm{max}$, computationally, represents the maximum volume fraction occupied by the central void in the hollow sphere, while in practice, this parameter represents the more general amount of deviation from a perfect homogeneous sphere \citep{2016A&A...585A..13M}.

Using the grain size and porosity as the only free parameters, we then compute a grid of models to estimate whether we can reproduce the parameterized phase function.
As a consequence of the high inclination of the disk and the employed inner mask not all scattering angles between 0 and 180 degrees can be sampled. 
To account for this, we use the best-fit parameters determined for the disk and the applied masks to estimate range of valid scattering angle. The result of this exercise is shown in Fig.~\ref{fig:scatter_ang}, yielding observable scattering angle from 9.14\degree\ to 170.86\degree\ ($\theta_{\mathrm{min,max}}=90\degree\pm80.86\degree$). 
We also note at this point that due to the method employed, the uncertainties in our disk model may be underestimated. We therefore utilize the 3$\sigma$ error bounds on the value of $g$, rather than the $1\sigma$ values.

\begin{figure*}[htbp] 
	\includegraphics[width=180mm]{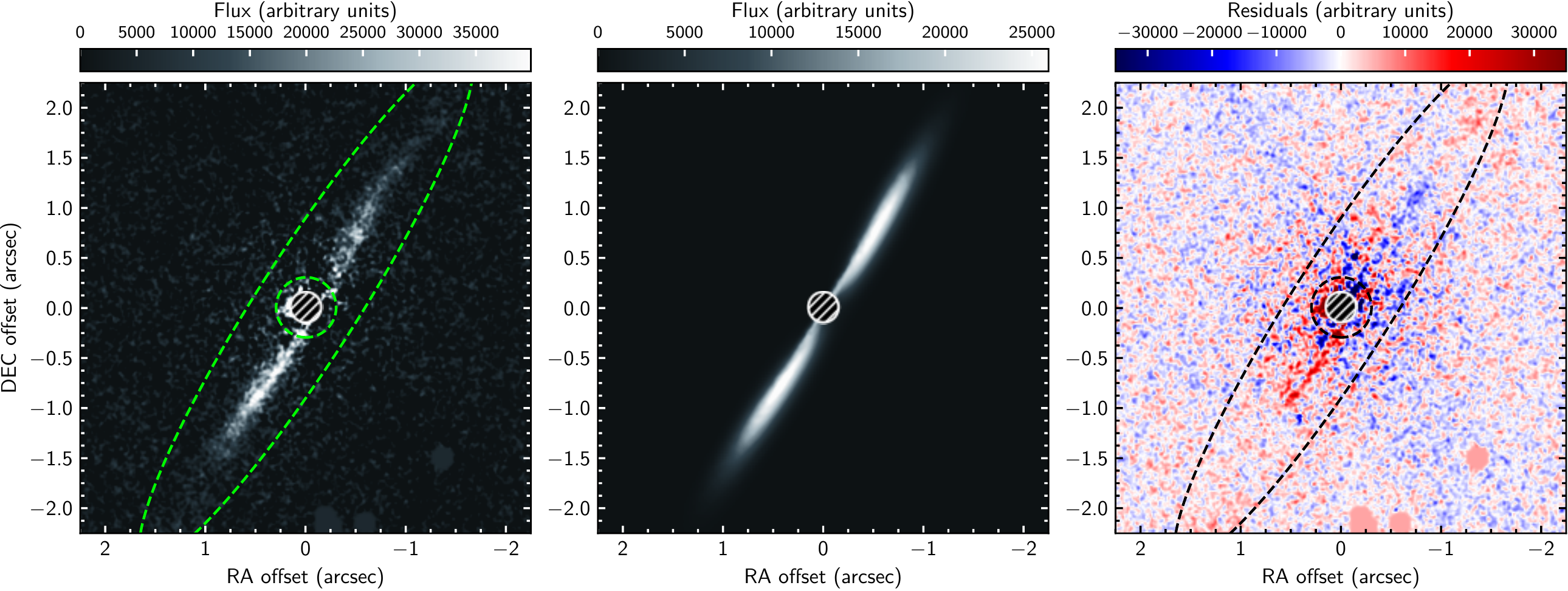}
	\caption{Observation and model images for \gsc. Shown from left to right are the measured $Q_{\phi}$, best-fit model and residual image obtained from our disk model using a linear scaling for each panel. The regions outside of the ellipse and within the circle (green dashed lines) are excluded from the $\chi^{2}$ calculation. The coronagraphic mask is indicated by the (shaded) circular region in each panel. North is to the top and east to left in each panel.}
	\label{fig:data_model}
\end{figure*}

\section{Results}\label{sec:results}

\subsection{Geometric properties of the disk}

\begin{table}[hbtp!] 
	\caption{Model details and best-fit parameter for the disk modeling of the SPHERE/IRDIS observations and the modeling of the polarized phase function.} 
	\label{table:3}	
	\centering	
	\begin{tabular}{l c c}
		\hline\hline\\[-1em]
		Parameter & Uniform prior & Best-fit value \\
		\hline 
		$r_{0}$\,[au]							& [10, 300]  	& 107$\pm$2 \\[0.75mm]
		$\delta_{r}$\,[au] 						& [0, 200] 	 	& 27$\pm$1 \\
		$i$\,[$^{\circ}$]  						& [60, 90] 		& 84.3$\pm$0.3 \\
		$\phi$\,[$^{\circ}$]					& [100, 180] 	& 148.7$\pm$0.7 \\
		$g$										& [0, 1] 	 	& 0.60$\pm$0.03 \\
		\hline 
		$s$\,[\microns]	                	   	& [0.01, 1000] 	& ${\sim}0.3-1$  \\
		porosity		  			 		   	& [0.01, 0.99] 	& n.d.  \\
		\hline 
	\end{tabular}
	\tablefoot{$r_{0}$: Reference radius and center of the radial distribution. $\delta_{r}$: Standard deviation of the radial distribution. $i$: Inclination. $\phi$: Position angle. $g$: Scattering efficiency paramater. $s$: Grain size. The porosity could not be determined (n.d.) due to a degeneracy in the model, between the minimum grain size and the porosity of the dust particles.}
\end{table}

We first used the model described in the previous section to reproduce the observed disk around \gsc, and to determine the most probable values for each of the five free parameters ($r_{0}$, $\delta_{r}$, $i$, $\phi$, and $g$). 
To assess the convergence and stability of the MCMC solution, we estimated the maximum autocorrelation length among all parameters, i.e. the average autocorrelation time, after each iteration. 
The fitting is considered converged when the number of iterations is larger than 100 times the average autocorrelation time and its changes, between subsequent iterations, are less than 1\%. 
In Appendix \ref{fig:t_autocorr} we show the evolution of the autocorrelation time as a function of the iteration step over the course of the modeling.
At the end of the modeling, the average autocorrelation time was 79 steps and the mean acceptance fraction \citep{1992StaSc...7..457G} for our best-fitting model was 0.47.
The best-fit model, along with the residuals and the observations, are presented with the same linear scaling in the center, right and left panels of Fig.~\ref{fig:data_model}, respectively. 
The most probable parameters and their uncertainties are summarized in Table~\ref{table:3}, while the probability density functions, plotted with $\mathtt{corner}$ package \citep{2016JOSS....1...24F}, are shown in Fig.~\ref{fig:corner}. 

We found that our observations are best described by an extended disk with a dust distribution centered at a radius $r_{0}\approx107\pm2$\,au, with a standard deviation of the radial distribution of $\delta_{r}\approx27\pm 1$\,au which is highly inclined at an inclination of $i\approx 84.3\,\degree\pm0.3$ and a position angle $\phi\approx148.7\,\degree\pm0.1$.
For the analytical polarized phase function we found a coefficient $g=0.60\pm0.03$. 

The uncertainties for the MCMC result parameters are estimated from the 0.16 and 0.84 quartiles using the $\mathtt{corner}$ package, and are also shown along with the projected posterior distributions in Fig.~\ref{fig:corner}.
These uncertainties, however, are most likely underestimated and should be taken with caution. 
This might be because we computed the goodness of the fit using a noise map derived from the standard deviation in concentric annulii in the $U_{\phi}$ image, a strategy commonly employed in direct imaging studies. 
We therefore may be underestimating the true uncertainties, resulting in larger $\chi^{2}$ values. This in turn might also have led to a narrower probability distribution as the Monte Carlo algorithm samples a smaller range of parameter values.

\subsection{Dust properties}

\begin{figure*}[htbp!] 
	\includegraphics[width=180mm]{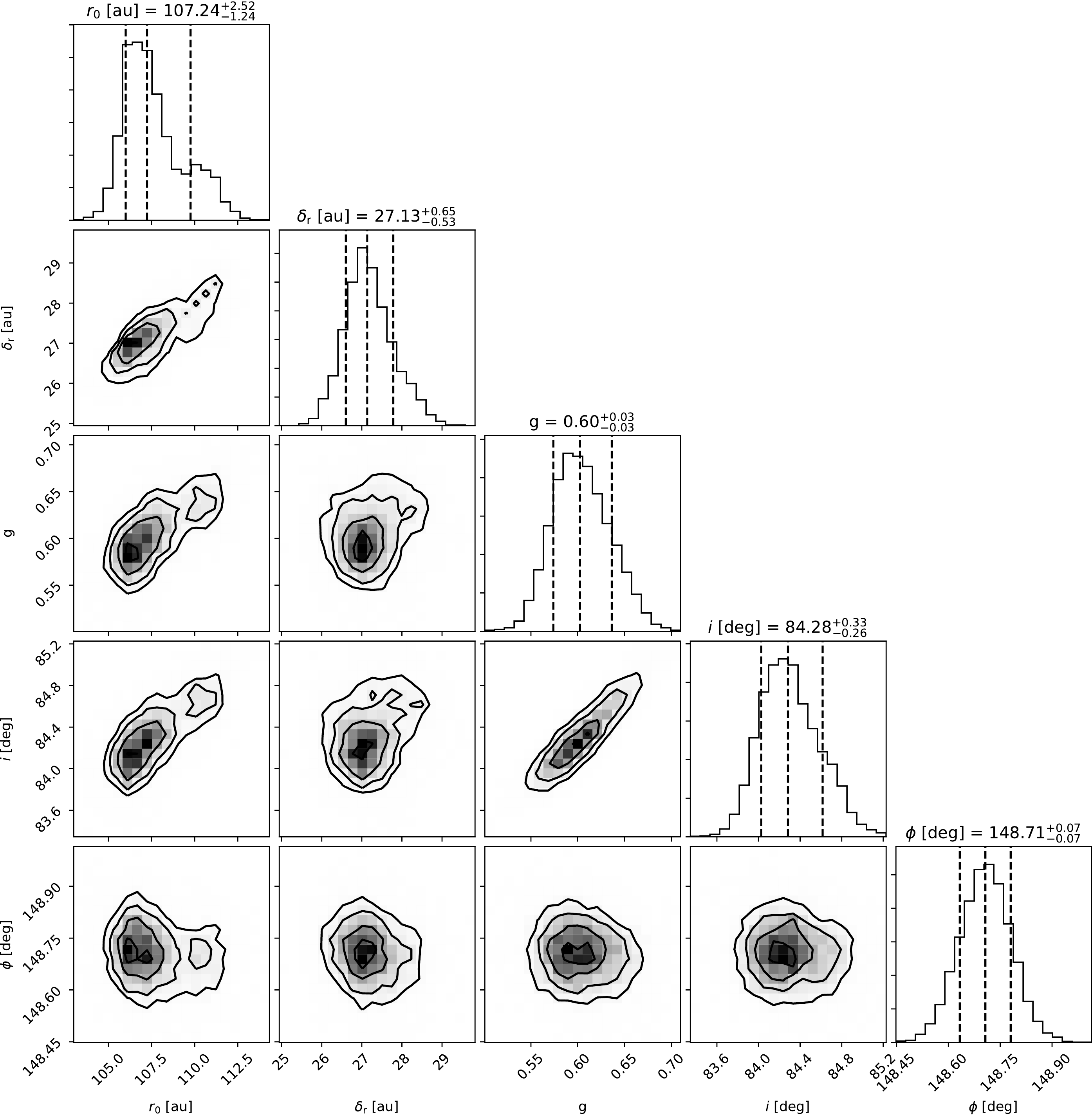}
	\caption{Projected posterior distribution and density plots of the free parameters used in the disk modeling. The plot additionally shows the 50\,\%, 16\,\%, and 84\,\% quartiles (vertical dashed lines), representing the distributions median and the 1\,$\sigma$ uncertainties (lower and upper bound), respectively. The 1D histograms represent the probability distributions of each parameter marginalized over the other free parameter. The contour maps represent the central 68.3\%, 95.5\%, and 99.73\% of the 2D probability distributions of different combinations of parameters, marginalized over each other.}
	\label{fig:corner}
\end{figure*}

\begin{figure}[hbtp!] 
	\begin{center}
		\resizebox{\hsize}{!}{\rotatebox{0}{\includegraphics{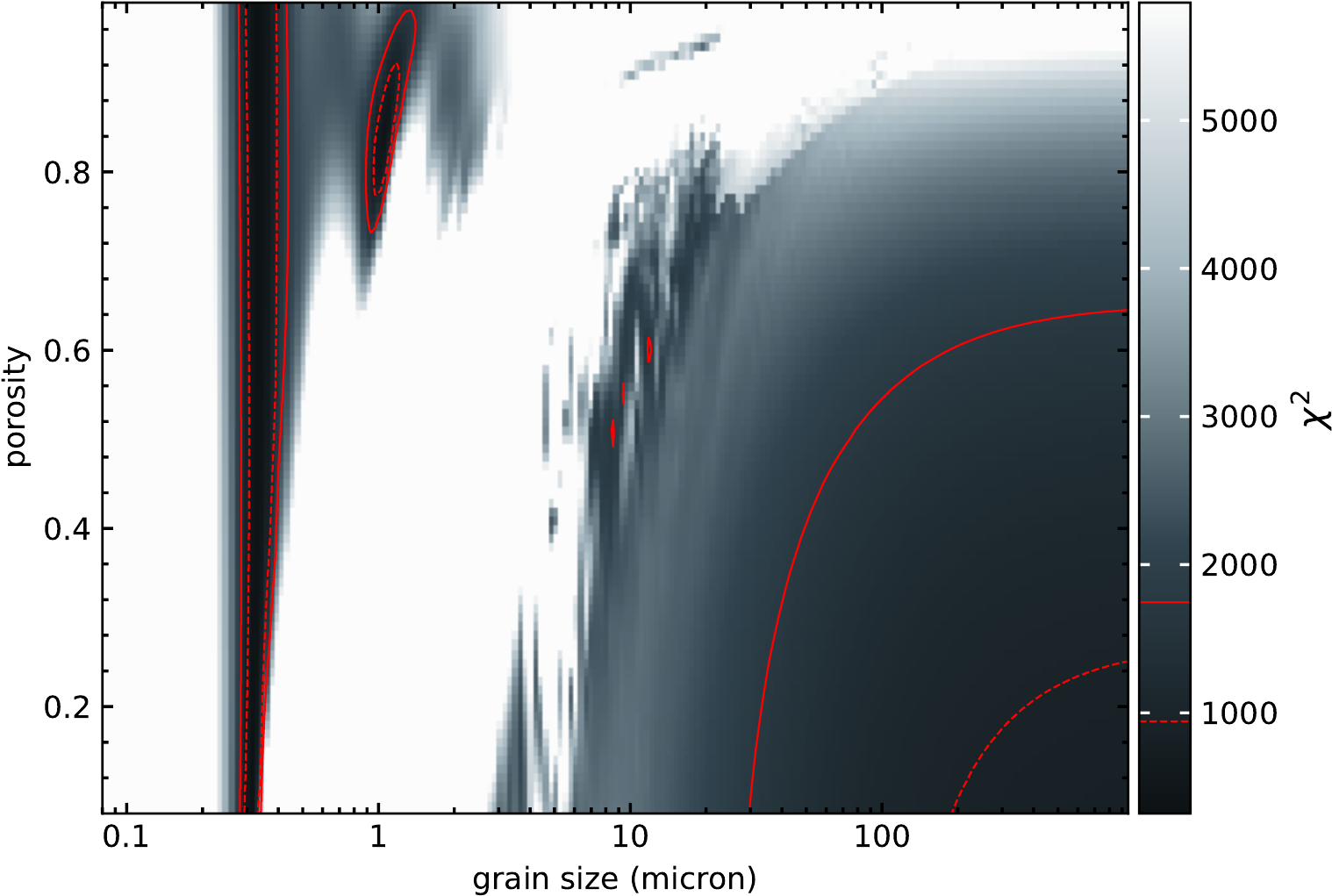}}}
	\end{center}
	\caption{2D $\chi^{2}$ map constructed from the computed $\chi^{2}$ at each point (model) of the parameter grid. The free parameter of the model are the porosity and the grain size. The contours enclose the regions where the $\chi^{2}$ is smaller than the 5\% (dashed line) and 15\% (solid line) quartiles. To increase the contrast the upper limit of the $\chi^{2}$ color map is set to the 50\% quartiles.}
	\label{fig:chi2_map}
\end{figure}

\begin{figure}[hbtp!] 
	\begin{center}
		\resizebox{\hsize}{!}{\rotatebox{0}{\includegraphics{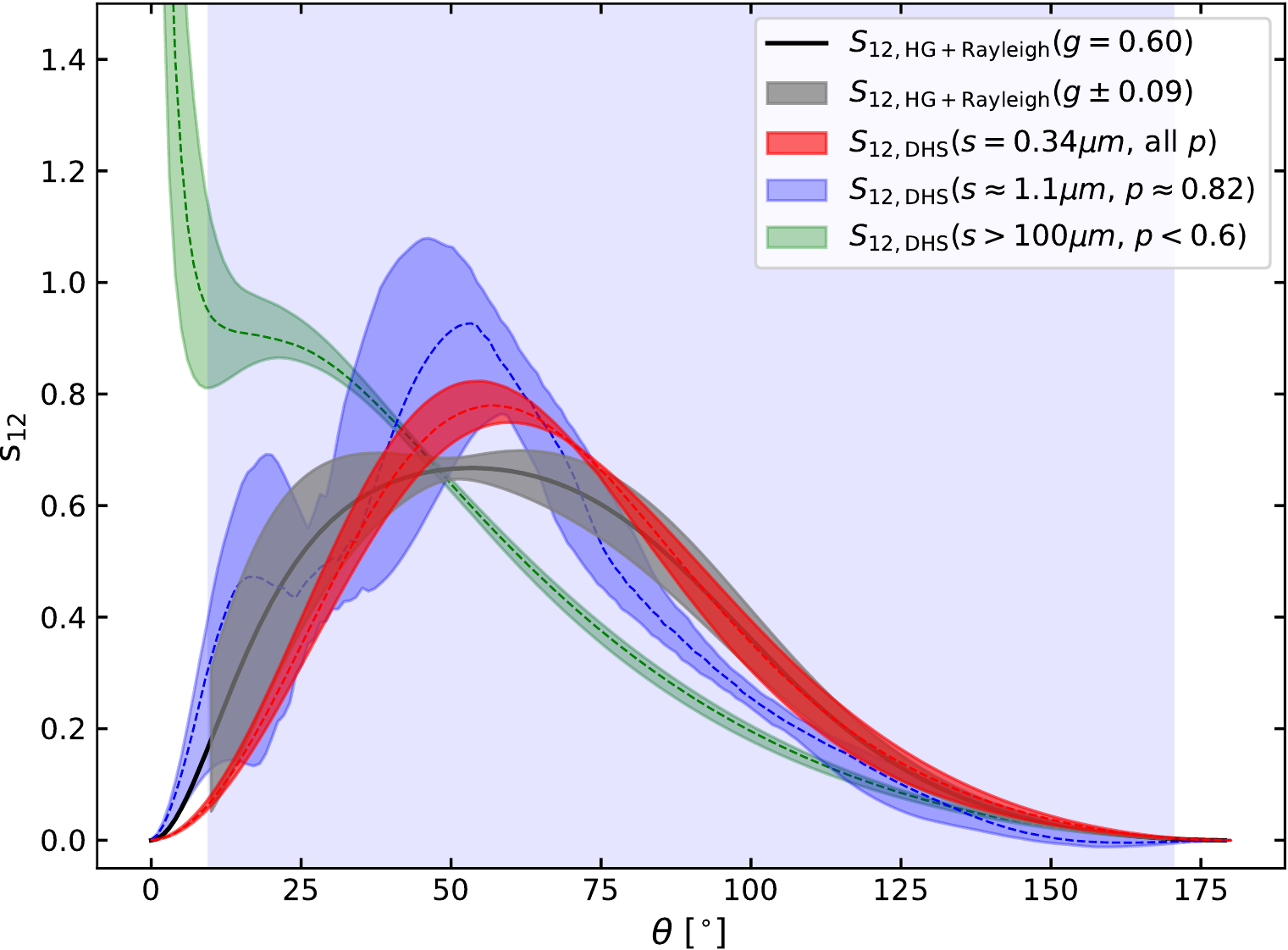}}}
	\end{center}
	\caption{Polarized phase function ($S_{\mathrm{12,\,HG}}$) as function of scattering angle $\theta$. Shown are the parameterized polarized phase function from our best-fit disk model for $g=0.60\pm0.09$ (black line and gray area) and the range (red area, enclosing values between the 16\% and 84\% quartiles) of best-fit model ($s=0.34$\,\microns, $\mathrm{porosity}=0.01-0.99$) of this phase function considering a distribution of hollow spheres. Additionally shown are the polarized phase functions resulting from the ``low''-$\chi^{2}$ regions indicated in Fig.~\ref{fig:chi2_map}. The phase functions were sampled around grain sizes and porosities of about 1\,\microns, and 0.80 (blue area), and grains larger than 100\,\microns\ and porosities below 0.6 (green area), respectively. The range of possible scattering angle ($\theta=9.14$\degree\ to $\theta=170.86$\degree) observable with the used mask is indicated by the blue shaded area.}
	\label{fig:phasefunc}
\end{figure}

\begin{figure}[hbtp!] 
	\begin{center}
		\resizebox{\hsize}{!}{\rotatebox{0}{\includegraphics{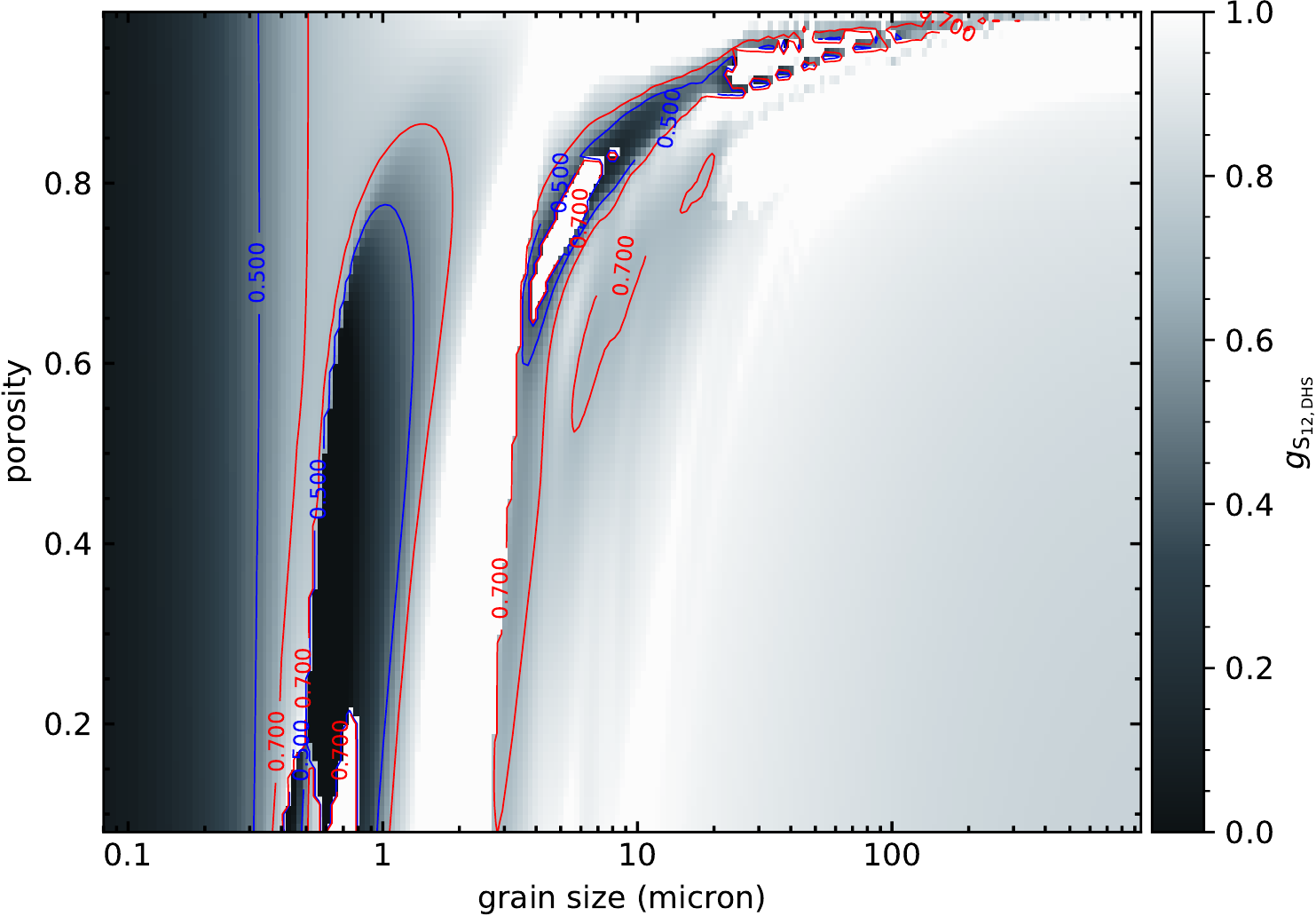}}}
	\end{center}
	\caption{2D map of the anisotropic scattering factor $g_{\mathrm{S_{12,\,DHS}}}$ computed at each point of the grain-size, porosity parameter grid with the DHS model. The contours for $g_{\mathrm{S_{12,\,HG}}}=0.50$ (blue lines) and $g_{\mathrm{S_{12,\,HG}}}=0.70$ (red lines) indicate the range of scattering factors from our best-fit disk model with the HG-approximation.}
	\label{fig:hg_map}
\end{figure}

\begin{figure*}[htbp!] 
	\centering
	\includegraphics[width=180mm]{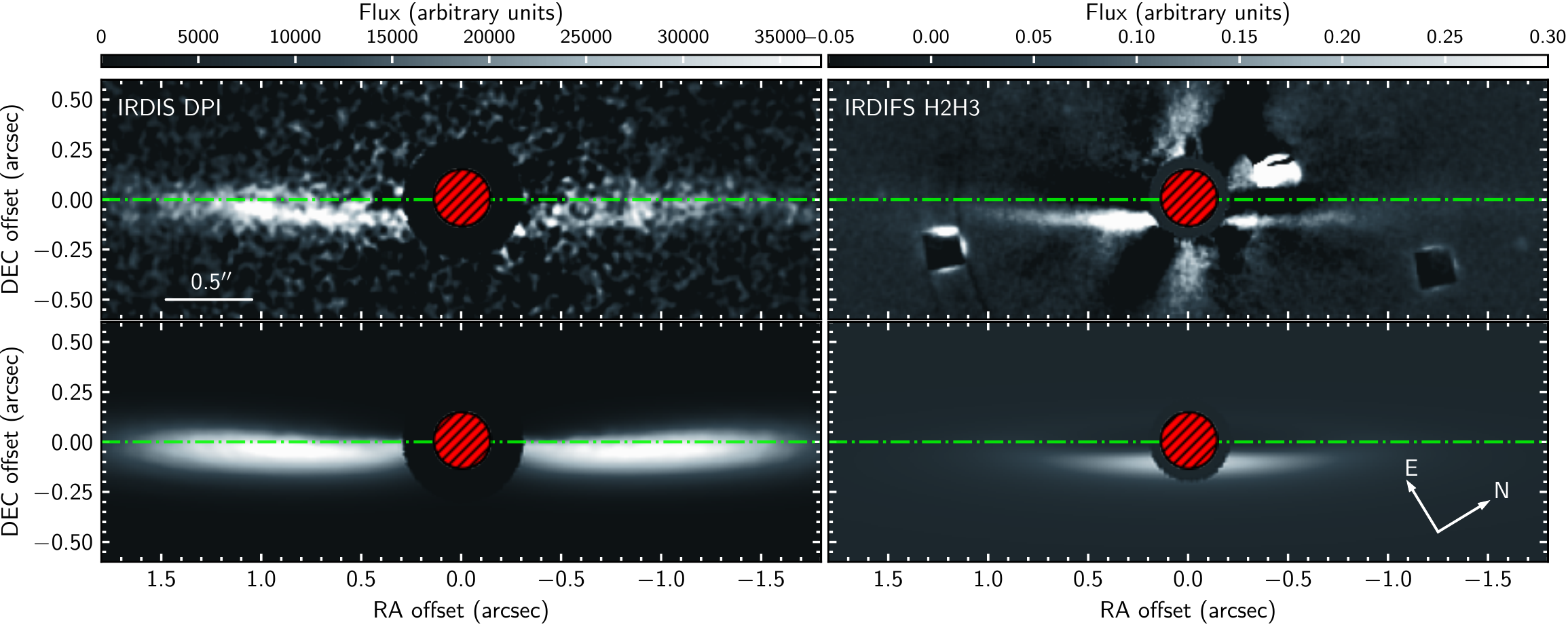}
	\caption{The disk around \gsc\ as seen with SPHERE/IRDIS in polarized light (top-left panel) and total intensity \citep[top-right panel,][]{2018A&A...613L...6S} in the H-band. The bottom panels show the best-fit disk model found in this work (bottom-left) and found by \citet{2018A&A...613L...6S} in the same scaling as the according observations.
	The dash-dotted line is parallel to the disk semi-major axis and crosses the central star. The coronagraphic mask is indicated by the (shaded) circular region in each panel.}
	\label{fig:comp_sissa}
\end{figure*}

\begin{figure*}[htbp!] 
	\centering
	\resizebox{0.9\hsize}{!}{\rotatebox{0}{\includegraphics{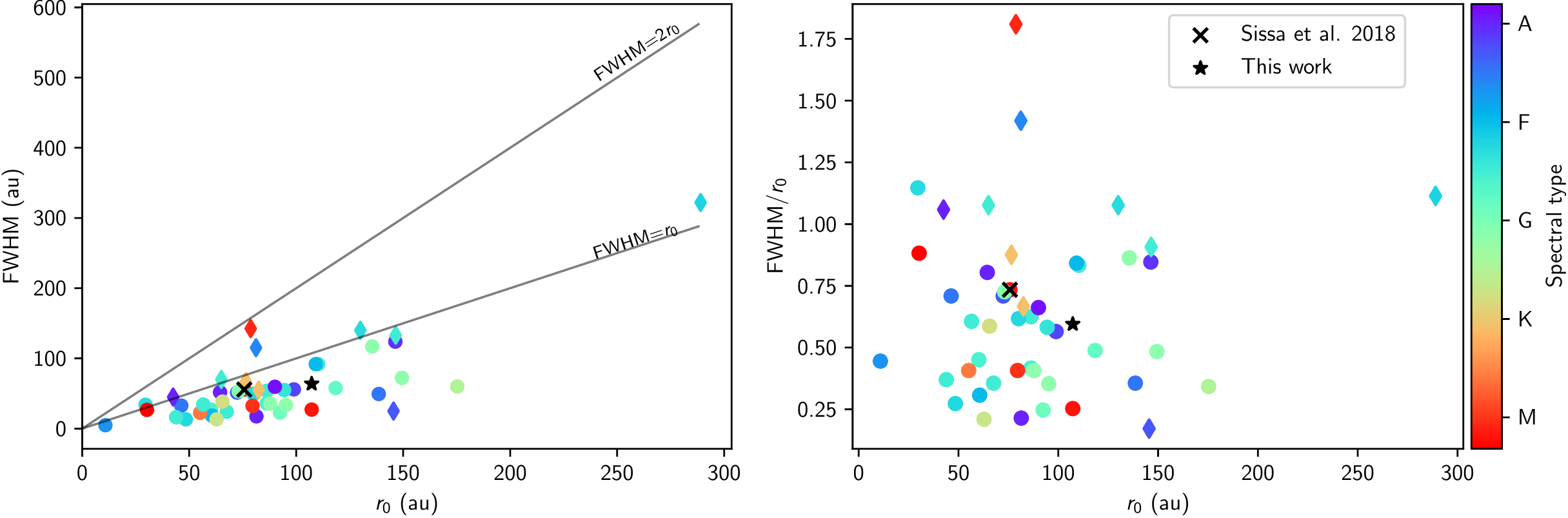}}}
	
	\caption{Full width at half maximum as a function of the center of the radial distribution $r_{0}$ (left panel), and FWHM normalized by the radius of the disk as function of $r_{0}$ (right panel). Objects for which the FWHM was estimated readily from a radial Gaussian or power-law distribution are shown as circles, while objects where only the inner and outer radius were given are marked by diamonds. The spectral types of the sample stars are indicated by the color bar. The radius and width of the disk around \gsc, estimated by \cite{2018A&A...613L...6S} and in this work, are marked by the cross and the asterisk, respectively. To guide the eye, we also show the limits for FWHM=$r_{0}$ and FWHM=2$r_{0}$ as gray line in the left panel.}
	\label{fig:fwhm}
\end{figure*}

\begin{figure}[hbtp!] 
	\begin{center}
		\resizebox{\hsize}{!}{\rotatebox{0}{\includegraphics{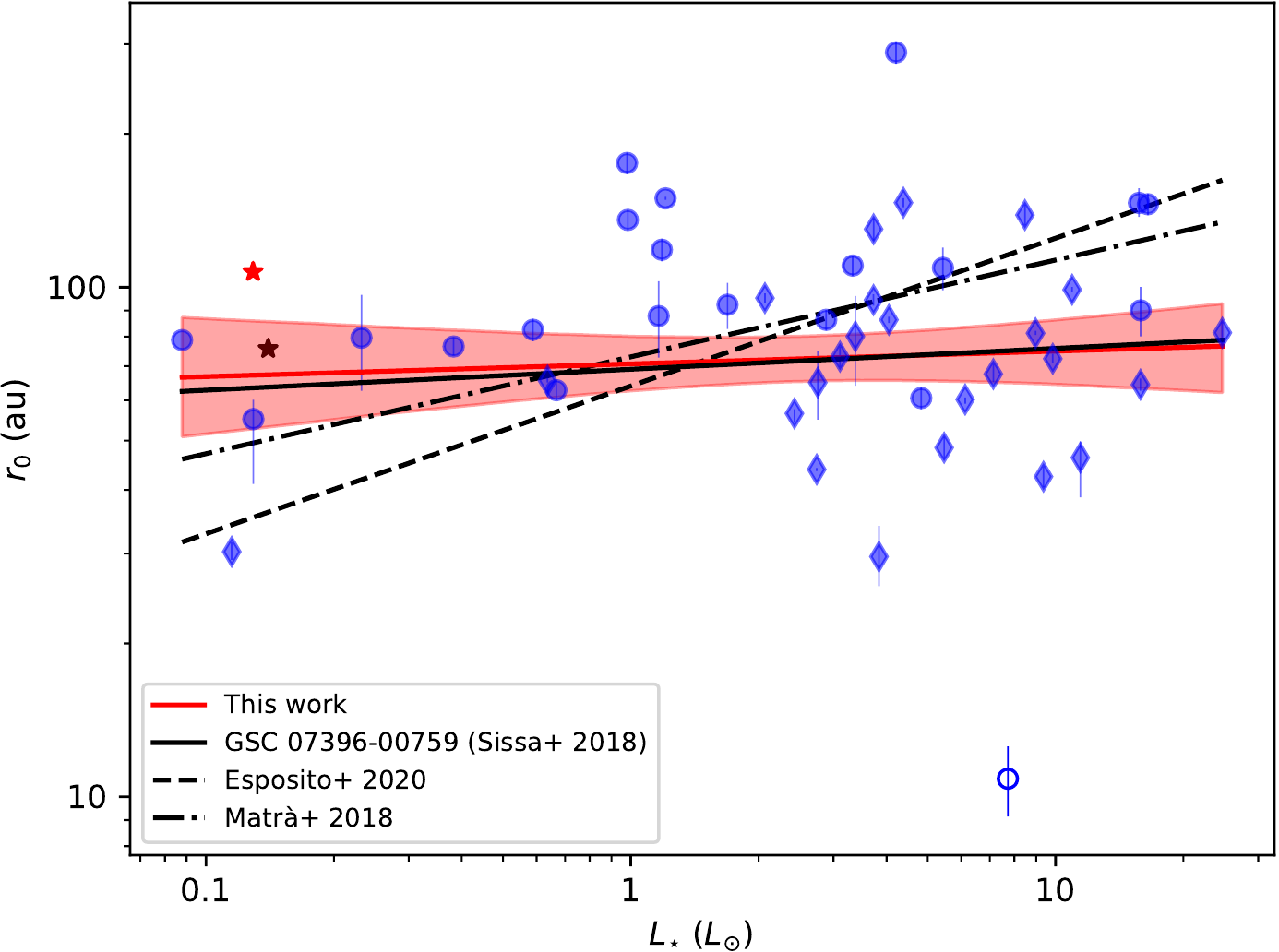}}}
	\end{center}
	\caption{Scattered light disk radii as a function of the stellar luminosity. Shown are the best-fit power-law function as solid (red) line and the 1-$\sigma$ confidence interval drawn from the corresponding probability distributions. Data points that are also part of the sample analyzed by \citet{2020AJ....160...24E} are marked as diamonds. Similar to \cite{2020AJ....160...24E}, we excluded HR\,7012 (blue open circle) from the fit as an outlier. The best-fit result using the radius estimate by \cite{2018A&A...613L...6S} (black star) is plotted as solid (black) line. Also shown for comparison are the radius–luminosity power laws for planetesimal belt central radii from thermal light imaging by \citet{2018ApJ...859...72M} (dash-dotted line), and the GPIES-detected disks in scattered light by \citet{2020AJ....160...24E} (dashed line), respectively.
}
	\label{fig:r-l}
\end{figure}

\begin{figure}[hbtp!] 
	\begin{center}
		\resizebox{\hsize}{!}{\rotatebox{0}{\includegraphics{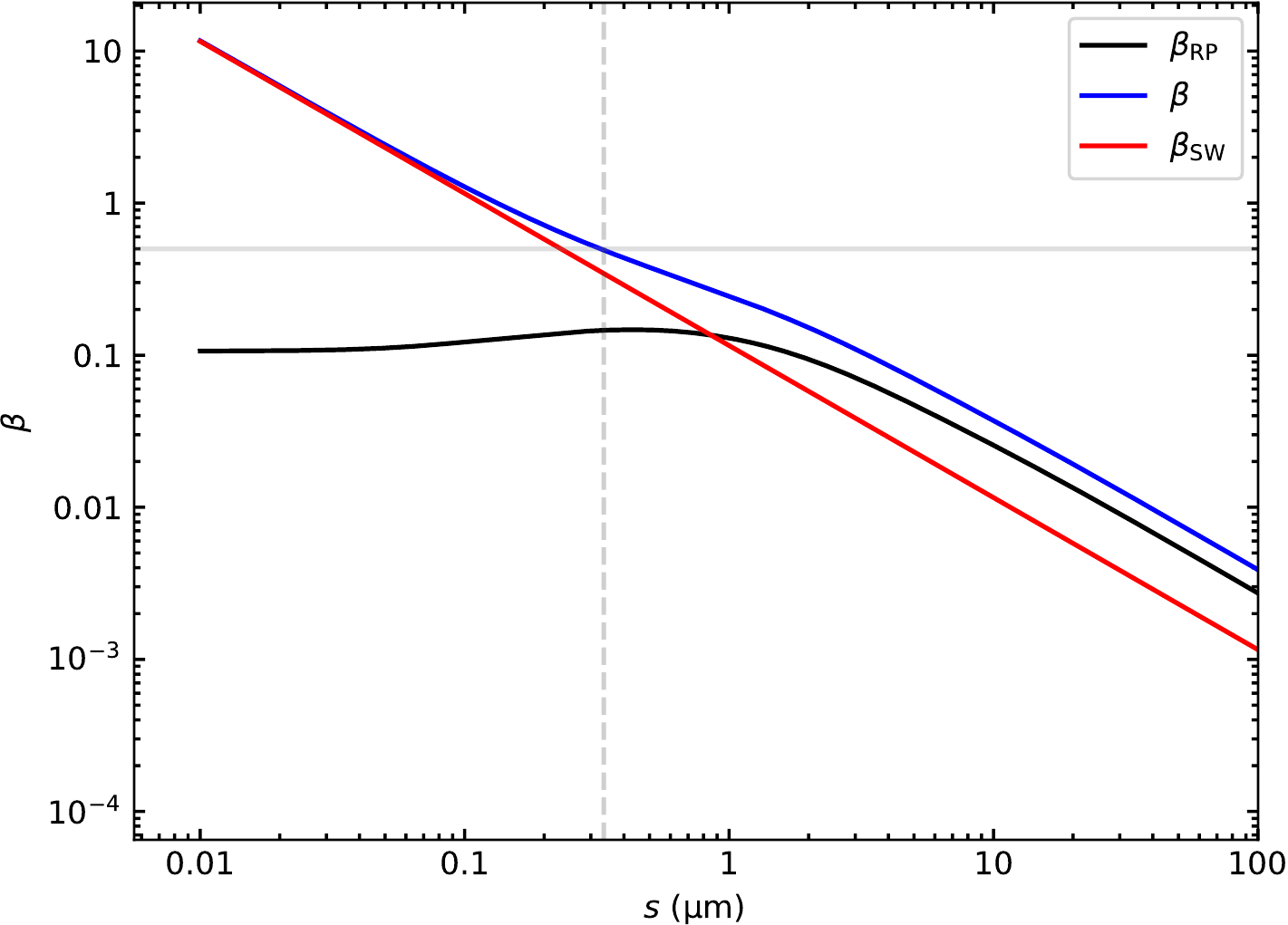}}}
	\end{center}
	\caption{$\beta$ as a function of grain size. $\beta_\mathrm{RP}$ and $\beta$ are shown as black and solid lines, respectively. The horizontal solid gray line ($\beta=0.5$) is the upper limit for bound trajectories assuming zero eccentricity for the parent body, see \cite{2017A&A...607A..65S} for further details. The vertical dashed line indicates the corresponding minimum grain size ($s=0.34\,\microns$) of dust particles required not to be expelled from the disk, assuming a porosity of 70\%.}
	\label{fig:beta}
\end{figure}

Next we applied a grid of models, computed with the $\mathtt{OpacityTool}$, as described in Sec.~\ref{subsec:pp_ml}, to estimate whether we can reproduce the inferred polarized phase function ($S_{\mathrm{12,\,HG}}$) considering a distribution of hollow spheres.
The resulting $\chi^{2}$ map is shown in Fig.~\ref{fig:chi2_map}.
Also shown are the contours that enclose the regions where the $\chi^{2}$ is smaller than the lower 5\% (dashed line) and 15\% (solid line) quartiles.
We found that the analytical form of the polarized phase function ($S_{\mathrm{12,\,HG}}$) is best reproduced by small micron-sized dust grains with $s{\sim}0.34$\,\microns. 
Due to a degeneracy between the grain size and the dust porosity, however, we cannot readily constrain the porosity for this grain size (see discussion below).

In Fig.~\ref{fig:phasefunc}, we present the parameterized polarized phase function ($S_{\mathrm{12,\,HG}}$) as a function of the scattering angle $\theta$ from our disk modeling for $g_{\mathrm{S_{12,\,HG}}}=0.60\pm0.09$ (black line and gray area), along with the best-fit model of the polarized phase function ($S_{\mathrm{12,\,DHS}}$, red line). Also shown is the 1\,$\sigma$ range of models (red area, enclosing values between the 16\% and 84\% quartiles of the likelihood distribution), estimated from the phase functions for $s=0.32-0.36$\,\microns, and the full range of modeled porosities (0.01-0.99). 
As can be seen, despite being the best solution to our model approach, there are notable differences between both curves, but in particular for scattering angle of less than 90\degree. 
The $\chi^{2}$ map, furthermore, not only shows "low" $\chi^{2}$ values around 0.34\,\microns, but also for particles of 1\,\microns\ with porosities around 0.8, and for larger particles ($s>100$\,\microns) with porosities of less than 0.6.
The corresponding range of phase functions, determined similar to the best-fit results, are shown as well in Fig.~\ref{fig:phasefunc}.
These phase functions clearly provide a worse fit to our approximation (shown in black) then the best solution using a DHS model (shown in red).

The presence of multiple solutions, on the other hand, could indicate additional hidden degeneracies beyond the one on the minimum size and porosity mentioned, restricted only to the second step of our alternative two step approach. 
For instance, we use a parameterized approximation ($S_{\mathrm{12,\,HG}}$) of the real phase function ($S_{\mathrm{12}}$) which effectively limits the examinable parameter space. 
We then fit the obtained curve with the phase function computed for a DHS model.
However, it is questionable whether we can reasonably reproduce the analytical HG approximation of the phase function with the phase function from a DHS model.

To address this question we created a map similar to the $\chi^{2}$ map shown in Fig.~\ref{fig:chi2_map}, where each point of the grid represents the best g value fitting the DHS model for a given set of grain size and porosity.
The resulting map of $g$ values (hereafter called $g_{\mathrm{S_{12,\,DHS}}}$) is shown in Fig.~\ref{fig:hg_map}. Also shown are the regions where the results overlap with the range of g values from the HG approximation. 
As can be seen, for the smaller grains (0.3--1\,\microns) the $g_{\mathrm{S_{12,\,DHS}}}$ value estimates are mostly consistent with the range of $g$ values using the HG approximation, while the $g$ values for the larger particles appear systematically higher. The data also indicate that there are possibly isolated solutions for particles with sizes between 5\,\microns\ and 10\,\microns. 
We also summarize these results in the lower part of Table~\ref{table:3}.

\section{Discussion}\label{sec:discussion}

\subsection{Disk properties}\label{subsec:disk}

Our results from the modeling of the dust distribution around \gsc\ indicate a radius of $\sim107$\,au for the radial distribution of the grain number density of the debris disk, and as shown in the residuals image in the right panel of Fig.~\ref{fig:data_model}, most of the signal coming from the disk has been successfully removed.

The modeling results are globally consistent with the results reported by \cite{2018A&A...613L...6S}.
In particular, we found that the estimated full width at half maximum (FWHM) of the disk of $2\sqrt{2\ln2}\,\delta_{r}\approx64$\,au ($\delta_{r}\approx27$\,au) as well as the inclination ($i\approx84\,\degree$) and position angle ($\theta\approx149\degree$) for the disk around \gsc\ are in good agreement with the results obtained by \citet[][$i\approx83\,\degree$, $\theta\approx149\degree$, and FWHM\,$\approx56$\,au]{2018A&A...613L...6S}, although their forward modeling of the angular differential imaging (ADI) observations of \gsc\ yielded a smaller disk radius ($r_{0}\approx70$\,au) than our model of the polarized light observations ($r_{0}\approx107$\,au). 

Our disk radius estimate is also slightly larger than the radius obtained by \cite{2018A&A...613L...6S} using a geometric approach, where the authors measured the position of the maximum brightness (disk spine) with respect to the apparent semi-major axis (i.e. the apparent disk radius obtained via forward modeling) as a function of the separation from the star after the removal of background stars in the five-mode PCA IRDIS image \citep[see Fig.~2,][]{2018A&A...613L...6S}.
Under the assumption that the disk is not flared, \cite{2018A&A...613L...6S} found that a ring of material with radius $r = 1.34\arcsec$ ($\sim 98$\,au) and inclination $i = 84.5\degree$ can properly describe the observed spine up to 1.2$\arcsec$ ($\sim 89$\,au) in their ADI images.
However, as also pointed out, the presence of at least three very bright stars may have altered the light distribution of the disk, and thus, the results have to be taken with caution as well.

It is surprising, nevertheless, that we found $r_{0}$ to be $\approx107$\,au ($\approx$\,1.5$\arcsec$) in our model, while we only detected uniform emission up to about $\approx70-80$\,au (1--1.1$\arcsec$) in the image. 
This suggests that according to our model, we do not fully recover the major axis of the disk with our observations, because we also found that the maximum of the polarized phase function (see Fig.~\ref{fig:phasefunc}) is shifted to smaller scattering angle ($\theta\approx55\,\degree$), while the phase function at the same time is relatively low at scattering angles of 90\,\degree.

We suggest that this is due to a degeneracy between the reference radius of the density distribution $r_{0}$ and the value of the anisotropic scattering factor $g$ for the phase function. 
We emphasize, however, that this degeneracy is solely the case for highly inclined disks, because significant azimuthal information is lost in this particular, nearly edge-on configuration \cite[see e.g. Fig.~8,][]{2017A&A...607A..90E}.

Our value of $g=0.60$ therefore might be too large. Nevertheless, we can exclude very small $g$ values, because, if $g$ were close to zero, the major axis of the disk would become much brighter than the minor axis, both due to the phase function peaking closer to 90 degrees and the column density that is larger at the major axis of the disk \cite[see e.g.][for discussion]{2020A&A...640A..12O}.
This, however, is not what we observe in the images, because we see a more or less uniform distribution of polarized light along the disk, as opposed to two distinct lobes, seen for example in the modeling of the disk around HD~61005 \citep{2016A&A...591A.108O}.
The observed degeneracy might also play a significant role for ADI images, since forward-scattering is even more pronounced in ADI observations, and therefore, the results of \cite{2018A&A...613L...6S} may also be impacted by this, as indicated by the side-by-side comparison of the two observations shown in Fig.~\ref{fig:comp_sissa}.

To explore the possibility that our radius for the disk around \gsc\ is overestimated, we re-run our grid of models with the reference radius fixed to $r_{0}=69.9$\,au, obtained by \cite{2018A&A...613L...6S} using forward modeling, to test whether we can find a solution that fits our observations.
The resulting best-fit model in this scenario is shown in Fig.~\ref{fig:data_model2}. In particular, we found a standard deviation of the radial distribution of $\delta_{r}\approx15.3\pm 0.2$\,au, an inclination of $i\approx 78.5\,\degree\pm0.3$ and a position angle $\phi\approx149.2\,\degree\pm0.1$.
For the analytical polarized phase function we found a coefficient $g=0.33$. 
Although the model converged on a solution, the results clearly show that we under-estimate the reference radius, as in particular the outer regions are not well reproduced, when using the results from the forward modeling of the ADI observation by \cite{2018A&A...613L...6S}.
On the other hand, the reference radius we inferred is in fact quite close to the radius of the disk of $r_{0}\approxeq98$\,au obtained by \cite{2018A&A...613L...6S} from their geometric measurements.
We thus suggest that the estimated reference radius ($r_{0}\approxeq107$\,au), while not definitive, is probably a better estimate than the one inferred by \cite{2018A&A...613L...6S} using forward modeling.
However, an independent confirmation of the disk radius is required to settle the discussion.
One possible way to alleviate this degeneracy and to resolve the discrepancies between the different measurements is to obtain high angular resolution observations at longer wavelengths with the Atacama Large Millimeter Array (ALMA, Cronin-Coltsmann, in prep.).
These observations are more sensitive to larger dust grains, and would therefore allow us to place stronger constraints on the shape parameters of the debris disk, and consequently allow us to more reliably estimate the shape of the phase function.

\subsection{Disk radii}

Another important diagnostic of debris disks alongside the phase function and the size distribution of their dust are the dimensions of the debris disks, i.e. their extent, and radii.
Believed to be the by-product of the planet formation process, the dust observed in debris disks is thought to be produced by the grinding down of larger bodies, the planetesimals.
In general, theory would suggest that the presence of a planetesimal belt, be it caused by failed growth to planets or enhanced planetesimal formation due to complex and dynamical interaction with already formed planets, should be related to distance to the central star \citep{2018ApJ...859...72M}. 
Regardless of the underlying processes, it is clear that the radii of debris disks contain valuable information on the formation processes of planetesimals and planets.

Several studies of the radius-luminosity (R--L) relation have been conducted over the years on this topic, such as for \textit{Herschel} PACS-resolved disk radii between 70\,\microns\ and 160\,\microns\ \citep{2014ApJ...792...65P}, for planetesimal belt central radii from thermal light imaging \citep{2018ApJ...859...72M}, and recently on 25 GPIES-detected debris disks ($\lstar= 2-14$\,\ls) in scattered light by \citet{2020AJ....160...24E}.
These studies found only a marginal correlation between the luminosity of the star and the radius of the disk, indicating that the morphology of debris disk is likely a result of temperature-independent processes, such as for instance, shaping of planetesimal populations by
planets, stirring by various mechanisms, and long-term collisional depletion \citep{2014ApJ...792...65P}.

Despite the interesting results, the studies lack stars with L<2\,\ls\ at the lower end of the mass spectrum which may have introduced a selection bias affecting their results. 
With the newly available scattered light data from resolved disks around low-mass stars with $\lstar<2$\,\ls, we therefore compiled a list of stars with resolved debris disks, and observed in scattered light from the literature, and attempt to test for a possible correlation between the size of the disk and the properties of the host star, such as its spectral type or luminosity.
We chose to primarily focus on scattered light observations, because the size and radius of the observed disk depends also on the wavelength in which they are observed, as well as the strength of radiation pressure and stellar winds depending on the evolutionary state of the star.

Our sample contains a total of 46 stars (see Table~\ref{table:4} in the appendix), including \gsc, with spectral types ranging from B9 to M3, ages between 10\,Myr and 2\,Gyr, and stellar luminosities reaching from 0.1\,\ls\ to 25\,\ls.
In case there are several publications for a given target, we focused on those publications with the highest S/N observations, and a similar modeling approach to ours.
We then used the published parameters that describe the radial dust density distribution of the disk, such as $r_0$, $\Delta r$, $\alpha_{\mathrm{in}}$, and $\alpha_{\mathrm{out}}$, to estimate the FWHM as a function of the ``peak'' radius of the radial distribution $r_0$ along the mid-plane of the disk.
Out of the total sample presented, for 11 stars only the detected inner and outer extent of the disk ($r_{\mathrm{min}}$, $r_{\mathrm{max}}$) were given. 
In these cases we equate the peak radius $r_{0}$ to the mean of $r_{\mathrm{min}}$, and $r_{\mathrm{max}}$, and the difference $r_{\mathrm{max}} - r_{\mathrm{min}}$ as width of the disk, respectively.
For the other 35 stars, the estimated FWHM is based on a radial distribution that was directly derived either from the standard deviation of the assumed Gaussian, or a radial distribution $R(r)$ that uses two power laws and is given by the expression:
\begin{eqnarray}
R(r) &=& \left[ \left( \frac{r}{r_{c}} \right)^{-2\alpha_{\mathrm{in}}} + \left( \frac{r}{r_{c}} \right)^{-2\alpha_{\mathrm{out}}} \right]^{-1/2} \, , \;
\end{eqnarray}
where $r$ is the radial coordinate in the equatorial plane, and $r_{c}$ is a critical radius that divides the ring into inner and outer regions with separate density power law indices of $\alpha_{\mathrm{in}}$ and $\alpha_{\mathrm{out}}$, respectively \citep{1999A&A...348..557A}.
As also pointed out by \citet{1999A&A...348..557A}, the maximum of the dust density does not occur at $r_{c}$ but at a peak radius calculated via
\begin{eqnarray}
r_0 &=& \left(\frac{\Gamma_{\mathrm{in}}}{-\Gamma_{\mathrm{out}}}\right)^{\left(2\Gamma_{\mathrm{in}} - 2\Gamma_{\mathrm{out}}\right)^{-1}} r_{c}\, , \;
\end{eqnarray}
where $\Gamma_{\mathrm{in}}=\alpha_{\mathrm{in}}+\gamma$ and $\Gamma_{\mathrm{out}}=\alpha_{\mathrm{out}}+\gamma$.
For this analysis we set $\gamma=1$, thus assuming the scale height is a constant fraction of the radius throughout the disk.

To estimate the dependency of the disk radii on the stellar luminosity we follow the description in \citet{2018ApJ...859...72M} and use a power-law model where the belt locations $R_{i}$ (in au) are linked to their host star luminosities $L_{\star,i}$ (in \ls) through the form $R_{i}=R_{1\mathrm{L}_{\odot}}L_{\star,i}^{\alpha}+\epsilon_{i}$. The parameter $\epsilon_{i}$ represents the intrinsic scatter of the distribution, and is assumed to follow a Gaussian distribution with standard deviation $\sigma_{\mathrm{intr}} = f_{\Delta R} R_{i}$. 
The best solution to this power-law model is determined using an affine invariant ensemble sample Monte-Carlo Markov Chain \citep[$\mathtt{emcee}$ package,][]{2013PASP..125..306F}. 
We then draw random samples from an uninformative uniform prior on the free parameters $R_{1\mathrm{L}_{\odot}}$, $\alpha$ and $f_{\Delta R}$, using a likelihood function described by Equation (24) in \citet{2007ApJ...665.1489K} to determine the posterior probability distributions for the parameters.
To facilitate the comparison with \citet{2020AJ....160...24E}, we excluded HR\,7012 from the fit as an outlier too.

In Fig.~\ref{fig:fwhm} and Fig.~\ref{fig:r-l} we present the results of this analysis. Figure~\ref{fig:fwhm} shows the estimated FWHM (left panel) and the FWHM divided by $r_0$ (right panel), respectively, as function of the radius of the disk $r_0$.
To point out that the sample is not homogeneous, we mark the stars where only the inner and outer radius of the disk is given with a diamond, while the results obtained from density distributions using a Gaussian or power laws to describe the radial profile are marked by circles.
The stars are color-coded according to their spectral type.
We additionally indicated the limits for FWHM=$r_{0}$ and FWHM=2$r_{0}$ to guide the eye.
The largest disk contained in our sample is that of HD~15745, a F2V star whose disk was discovered by \cite{2007ApJ...671L.161K}. 
The star is one of 10 stars that have a disk with a FWHM larger than the radius of the respective disk, with the largest width-to-radius ratios being found for AU~Mic \citep{2005AJ....129.1008K}, HD~15745, and $\beta$~Pic \citep{2015ApJ...811...18M}.
Most of the stars have FWHM smaller than the radius of the disk, and overall we found neither an obvious correlation, nor clustering of data, between the radius of the disk, its radial extent, and the spectral type of the star.

From the analysis of the radius--luminosity relationship in our sample we found $R_{1\mathrm{L}_{\odot}} = 70.7^{+9.6}_{-8.3}$\,au, $\alpha= 0.02\pm0.08$, and $f_{\Delta R}= 0.07\pm0.02$, taking the $50_{-34}^{+34}$ percentiles of the posterior probability distributions for each parameter.
The results are illustrated in Fig.~\ref{fig:r-l}. Shown are the scattered light disk radii as a function of the stellar luminosity of our sample stars, as well as our best-fit power-law function as solid (red) line and the 1-$\sigma$ confidence interval estimated from a randomly drawn sample of the corresponding probability distributions shown in Fig.~\ref{fig:corner_r-l}.
The data points that are also contained in the sample analyzed by \citet{2020AJ....160...24E} are marked as diamonds. 
Because of the inconsistent radius estimates between this work and the work of \cite{2018A&A...613L...6S}, we repeated the fit using the radius estimate by \cite{2018A&A...613L...6S} and show the result as solid (black) line. However, it is apparent that the derived relation is not strongly dependent on which estimate of the radius of the disk around \gsc\ is included.
For comparison, we furthermore plotted the radius–luminosity power laws for the planetesimal belt central radii from thermal light imaging by \citet[][$\alpha = 0.19\pm0.04$, dash-dotted line]{2018ApJ...859...72M}, and the GPIES-detected disks in scattered light by \citet[][$\alpha = 0.25\pm0.09$, dashed line]{2020AJ....160...24E}, respectively.

As can be seen, our result differs from the results reported by \citet{2018ApJ...859...72M} and \citet{2020AJ....160...24E}, both reporting evidence of a statistical relation between the radius and the luminosity (although marginal), while our result suggest that there is no correlation between the stellar luminosity \lstar\ and the scattered-light radius $r_{0}$. This would be consistent with the \citet{2014ApJ...792...65P} finding of no significant correlation between \lstar\ and $r_{0}$ detected in the Herschel PACS survey, and further support for the idea that the dimensions of debris disks are likely set or influenced by temperature independent processes like collisions, or dynamical interaction with planetary perturber.
However, we want to emphasize that most of the Herschel detections are marginally resolved and therefore are not directly comparable to our results. The presented sample was also derived from different observations, with varying spatial resolution, contrast performance at short angular separations, sensitivities, and differing modeling strategies. 
One should furthermore note that, as it is the case for \gsc, most of the disks resolved, in particular around low-mass stars, are highly inclined, making them favorable for detection in scattered light, but also susceptible to projection effects and inaccurately estimated phase functions.
The influence of stellar winds on the shape of the disk, in particular on the radial distribution of the smaller dust grains, also remains rather unclear due to the low number of resolved disks around low mass stars ($\lstar < 1\,\ls$). 
It is thus also possible that some radii are over- or under-estimated and may not necessarily measure the “peak” radius of the disk (i.e., the radius of the dust-producing planetesimal belt).

Because of these potential sources of uncertainty in the scattered-light radius measurements, the lack of correlation should be taken with caution, and better constraints on the radii of resolved disk around low mass stars using independent observation methods are required to better understand the relation between the luminosity of the star, the peak of the radial dust distribution of debris disks observed in scattered-light, and its implications for the formation and evolution of the dust-producing parent belt.

\subsection{Dust properties}\label{subsec:dust}

Despite the aforementioned assumptions and caveats about our modeling strategy, the best fitting model does account for most of the observed signal, the probability density functions of their parameters appear well constrained, and the estimated disk properties are compatible with the previous result by \citet{2018A&A...613L...6S}. 
We therefore used the inferred parameterized phase function to further investigate the dust properties of the disk around \gsc, as described in Sec.~\ref{subsec:pp_ml}.

The $\chi^{2}$ map obtained from our fit of the parameterized phase function is presented in Fig.~\ref{fig:chi2_map}. 
We found that the parameterized polarized phase function is relatively well reproduced by small dust grains of 0.34\,\microns\ (see Fig.~\ref{fig:phasefunc}).
The plot however also revealed a degeneracy between the grain size and the porosity, that suggests that we are most likely only tracing the small dust particles (0.34\,--1\,\microns) that are dispersed throughout the disk. Composed of sub-\microns\ sized monomers these particles can also make up larger aggregates. 
As explained in \cite{2016A&A...585A..13M}, the polarization properties of such aggregates are intimately related to the size of the individual monomers and not to the overall size of the aggregate itself.
This degeneracy suggests that we cannot fully reconcile all key aspects (e.g., phase function, porosity and grain size) with a single scattering theory, a well known problem in the study of debris disks (see e.g. the review by \citealp{2018ARA&A..56..541H}, or the studies of HR~4796~A, or HD~191089 by \citealp{2017A&A...599A.108M,2019A&A...626A..54M,2019A&A...630A.142O}, and \citealp{2019ApJ...882...64R}, respectively). 
Since the dust grains can be aggregates composed of smaller particles, the inferred dust particle size of $s\approx0.34$\,\microns\ is most likely a lower limit for the dust grain size in the disk of \gsc.

Nevertheless, we can use this information to try to constrain the strength of the stellar winds.
To put very rough constraints on the mass-loss rate of \gsc, we used Eq.~\ref{eq:b_sw} and Eq.~\ref{eq:b_pr} described in Sect.~\ref{subsec:rtm} to compute the total net pressure force acting on a grain \citep[$\beta=\beta_\mathrm{RP}+\beta_\mathrm{SW}$,][]{2017A&A...607A..65S}, depending on the estimated minimum grain size $s_{\mathrm{min}}$, the blow-out size of the dust grains, for which $\beta \leq 0.5$ (assuming the parent bodies are on circular orbit). Consequently, all the grains that are smaller are no longer bound to the star, and thus would be removed from the system rapidly.
$Q_{\mathrm{RP}}$, the radiation pressure efficiency averaged over the stellar spectrum in Eq.~\ref{eq:b_pr}, was calculated using the asymmetry parameter $g_{\mathrm{sca}}$ which can also be computed with the $\mathtt{OpacityTool}$, and is equal to $Q_{\mathrm{ext}} (\lambda, s) - g_{\mathrm{sca}} (s) \times Q_{\mathrm{sca}} (\lambda, s)$.

Assuming a stellar luminosity $\lstar=0.13$\,\ls, a mass of $\mstar=0.62$\,\msun, a minimum blow-out size of 0.34\,\microns\ for the dust grains, we estimate that the requirement of having a sufficiently strong stellar wind is obtained for average stellar mass-loss rates ranging from $\mloss\approx10$\,\mlosssun\ up to $\mloss\approx500$\,\mlosssun, depending on the assumed porosity of the dust particles with this size.
For completeness we show in Fig.~\ref{fig:beta} the total net pressure force $\beta$ and the force exerted by the radiation pressure $\beta_{\mathrm{RP}}$ as a function of the grain size, exemplary for an assumed minimum blow-out size of $s_{\mathrm{min}}=0.34$\,\microns, and a porosity of 70\%, yielding a mass-loss rate of \mloss$\approx250$\,\mlosssun.
Although our results are consistent with findings for similar-type stars such as AU~Mic \citep[see e.g.][]{2015A&A...581A..97S,2017A&A...607A..65S}, with mass-loss rate between $\mloss\approx50$\,\mlosssun\ and $\mloss\approx300$\,\mlosssun\ for grain sizes ranging from 0.04\,\microns\ up to 0.34\,\microns, respectively, we cannot derive a definitive conclusion due to the degeneracies in our modeling approach.

\section{Conclusion} \label{sec:concl}

In this paper we presented high angular resolution polarimetric observations of \gsc\ obtained with SPHERE/IRDIS in the near-infrared employing a broadband $H$ filter. 
We have derived the stellar parameters from the SED, reconstructed using photometric data ranging from the UV to the MIR wavelengths. From the position in the Hertzsprung-Russell diagram (HRD) combined with theoretical isochrones and mass tracks \citep{2015A&A...577A..42B} we estimated an upper limit for the age of $\lesssim$\,20\,Myr, a stellar mass of ${\sim}0.62\,{\msun}$, and a stellar radius of $\rstar\sim0.71\,\rsun$.

With the newly determined stellar parameters, we then characterized the disk structure and modeled the dust distribution of the disk around \gsc\ using a radiative transfer model that takes into account the effects of radiation and stellar wind pressure which is likely more efficient in low-mass stars such as \gsc.
We found that the polarized light observation is best described by our model by an extended disk with a dust distribution centered at a radius $r_{0}\approx107\pm2$\,au, with a standard deviation of the radial distribution of $\delta_{r}\approx27\pm 1$\,au which is highly inclined at an angle $i\approx 84.3\,\degree\pm0.3$ and a position angle $\phi\approx148.7\,\degree\pm0.1$.
The derived scattering asymmetry parameter is $g=0.60\pm0.03$, and was estimated using a modified Henyey-Greenstein function that accounts for the angle-dependence of the fractional polarization, produced by the particle scattering, by adopting the Rayleigh scattering function as approximation.

We were not able to fully recover the major axis of the disk, because of its nearly edge-on configuration we observe in the almost uniform emission along the disk, and measure in the polarized phase function.
The anisotropic scattering factor $g$ and consequently the radius of the disk therefore may be overestimated, although we argue that the observations are reasonably well reproduced by our model, and the geometric structure of the disk seen in polarized light is consistent with total intensity images taken with SPHERE-IRDIS by \cite{2018A&A...613L...6S}.

We further studied the dust properties such as the shape and size of the dust particles using a polarized phase function model that we calculated with the $\mathtt{OpacityTool}$, and assuming a distribution of hollow spheres with a single grain size to fit the inferred parameterized polarized phase function.
We found that the polarized phase function is reasonably well reproduced by small micron-sized dust grains with $s\approx0.34$\,\microns.
However, the modeling results also imply that the DPI observations may only trace the small monomers that are part of larger aggregates \citep{2016A&A...585A..13M}, if they are present, and therefore the deduced dust grain size may only be a lower limit on the size of the particles dispersed throughout the disk.
The results, therefore, suggest that we cannot reconcile all key aspects of the disk using a single scattering theory such as the DHS theory to explain for example the shape of the phase function, or its dependence on the dust grain size and porosity. This is a long standing problem in the analysis of debris disks and further observations are required to solve this issue.
Such observations could be obtained for example using SPHERE/IRDIS, with the star-hopping mode, which would allow to retrieve both the polarized and scattered light phase function, and therefore the degree of polarization, which contains crucial information on the shape of the phase function, but is typically challenging to measure.

Nonetheless, the observed extent of the disk around \gsc\ is comparable to the width of other disks around low-mass stars such as AU~Mic or GJ~581, and appear less well constrained than the dust belts resolved in scattered light around higher mass stars such as HR~4796~A. 
However, the comparison of the width as a function of the radius of the disk for 46 stars of spectral type A to M-type and with resolved disks in scattered light to a first approximation does not show any significant correlation between the extent of the disk, its radius, and the spectral type of the star.
As a second diagnostic we also used the estimated radii, and analyzed their dependency on the luminosity of the host star. 
From our sample of 44 disk-bearing stars with luminosities ranging from 0.1\,\ls\ to 25\,\ls, we found that the $R-\lstar$ relation is best explained by a power-law function of the form $R=71^{+10}_{-8}\lstar ^{0.02\pm0.08}$.
This is consistent with no statistically significant correlation between belt radii, where most of the observed dust is released, and host star luminosities and lends further support to the idea that the dimensions of debris disks are likely set or influenced by temperature independent processes like collisions, or dynamical interaction with planetary perturber.
However, the peak distribution of the dust particle, and by extension the location of the planetesimal belt from which they originate, observed in scattered light may be overestimated due to projection effects, inaccuracies in the estimated phase function or in case of young low mass stars, like \gsc, the influence of the stellar activity, especially stellar winds. 
Thus, further studies are needed to better understand how and why the planetesimals belts arise, in particular around low mass stars where the distribution of small dust particle may be significantly affected by stellar winds, present in young, low mass stars.

Overall, we conclude that if the disk around \gsc\ is indeed this extended ($\delta_{r}\approx30$\,au), and the disk is dominated by small dust grains of 0.34\,\microns, then the stellar winds could be as strong as 500 times the solar mass loss rate in \gsc, and, thus, could play a dominant role in the transport of particle into the outer disk that would otherwise remain closer to their parent bodies.
We note, however, that the mass-loss rates estimates should be taken with caution, because of the observed degeneracies in our models. 
Furthermore, our mass-loss rate primarily is an averaged one over time, because the considerations above do not take into account that $\beta_{\mathrm{SW}}$ for a given dust grain size may be time-variable because of fluctuations in the stellar flux caused by intense and frequent flares. This is common for young and low-mass stars such as \gsc, although the light curves of \gsc\ obtained with TESS and during the ASAS-SN survey do not show signs of stellar flares.
Nevertheless the coronal activity of the star suggested by the observed variations in the UV and X-ray, may not be negligible, and, thus, should be taken into account in future studies in order to better understand the influence of the different pressure forces acting on the dust grains and that are responsible for the observed morphology of the disk.

\begin{acknowledgements}
	We are extremely grateful to the anonymous referee for carefully reading the manuscript, and providing very helpful comments to improve the quality of our manuscript. C.\,Adam, J.\,O., A.\,B., and M.\,M. acknowledge support by ANID, -- Millennium Science Initiative Program -- NCN19\_171. J.\,O. acknowledges financial support from Fondecyt (grant 1180395). A.\,B. acknowledges financial support from Fondecyt (grant 1190748). A.\,Z. acknowledges support from the FONDECYT Iniciaci\'on en investigaci\'on project number 11190837.
	T.\,H. acknowledges support from the European Research Council under the Horizon 2020 Framework Program via the ERC Advanced Grant Origins 83 24 28.
	This research has made use of the SIMBAD database, operated at CDS, Strasbourg, France. This work has made use of data from the European Space Agency (ESA) mission
	{\it Gaia} (\url{https://www.cosmos.esa.int/gaia}), processed by the {\it Gaia} Data Processing and Analysis Consortium (DPAC, \url{https://www.cosmos.esa.int/web/gaia/dpac/consortium}). Funding for the DPAC has been provided by national institutions, in particular the institutions participating in the {\it Gaia} Multilateral Agreement. 
	This publication makes use of VOSA, developed under the Spanish Virtual Observatory project supported by the Spanish MINECO through grant AyA2017-84089.
	VOSA has been partially updated by using funding from the European Union's Horizon 2020 Research and Innovation Programme, under Grant Agreement nº 776403 (EXOPLANETS-A).
	This paper includes data collected by the TESS mission, which are publicly available from the Mikulski Archive for Space Telescopes (MAST). Funding for the TESS mission is provided by the NASA Explorer Program. 
	This research made use of Lightkurve, a Python package for Kepler and TESS data analysis (Lightkurve Collaboration, 2018). 
\end{acknowledgements}

%
%

\bibliographystyle{aa}
\bibliography{GSC07396-00759_adam_references.bib}

\clearpage
\begin{appendix} 

\section{Light curves}

To measure the stellar rotation period of \gsc\ we analyze available light-curves of \gsc\ observed with the Transiting Exoplanet Survey Satellite \citep[TESS,][]{2015JATIS...1a4003R}, and during the All-Sky Automated Survey for Supernovae \citep[ASAS-SN,][]{2014ApJ...788...48S,2017PASP..129j4502K}, respectively. \gsc\ was observed with TESS (600--1000\,nm), covering an observation period of about 27 days, between 2019-06-19 and 2019-07-17 and shown in Fig.~\ref{fig:lc_tess}.
The ASAS-SN light curve, on the other hand, was obtained in the $V$-band over a time period of 2.5 years between 2016-03-10 and 2018-09-22 and is presented in Fig.~\ref{fig:lc_asas-sn}. The period estimate was obtained using Lightkurve, a Python package for Kepler and TESS data analysis \citep{2018ascl.soft12013L}. After each light curve was pre-processed, i.e. outlier removal and normalization was applied, we determine the stellar rotation period $P_{\mathrm{rot}}$ from the corresponding Lomb–Scargle periodogram \citep{1976Ap&SS..39..447L,1982ApJ...263..835S}. 
The uncertainty associated with each period was evaluated by re-sampling the light curve using the bootstrap method (with replacement), and corresponds to the 95\% confidence interval of our sample estimates. It should be noted, however, that these are only statistical uncertainties and they do not represent the totality of the error budget of the light curves. 
From the TESS light curve of \gsc, we estimate a stellar rotation period $P_{\mathrm{rot}}=11.63\pm0.02$\,d, whereas the ASAS-SN light curve yielded a period $P_{\mathrm{rot}}=12.06\pm0.02$\,d, respectively.

\begin{figure*}[h] 
\includegraphics[width=18cm]{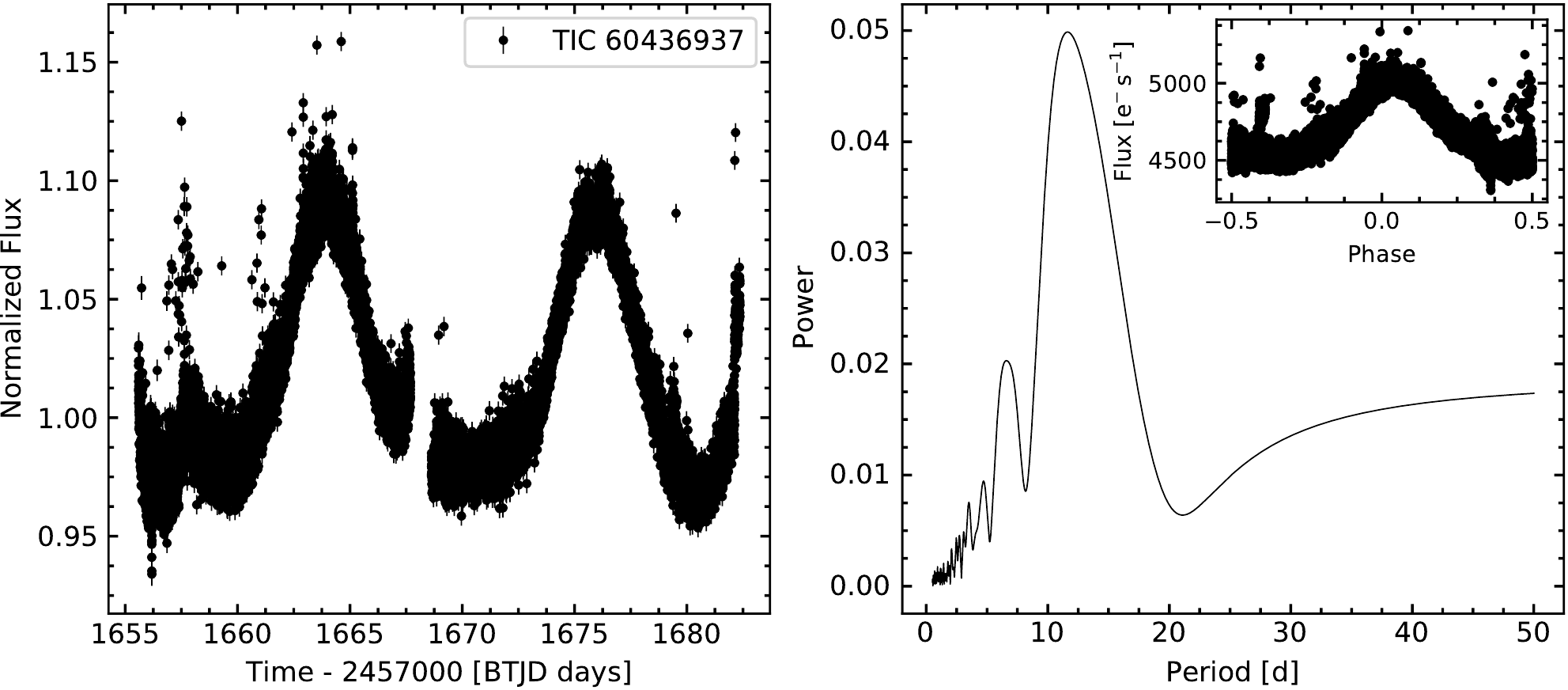}
\caption{Left panel: Normalized TESS light curve of \gsc, observed between 2019-06-19 and 2019-07-17. Right panel: Lomb-Scargle periodogram computed from these data, with an inset showing the light curve folded over the detected stellar rotation period $P_{\mathrm{rot}}=11.63$\,d.}
\label{fig:lc_tess}
\end{figure*}
\begin{figure*}[h] 
\includegraphics[width=18cm]{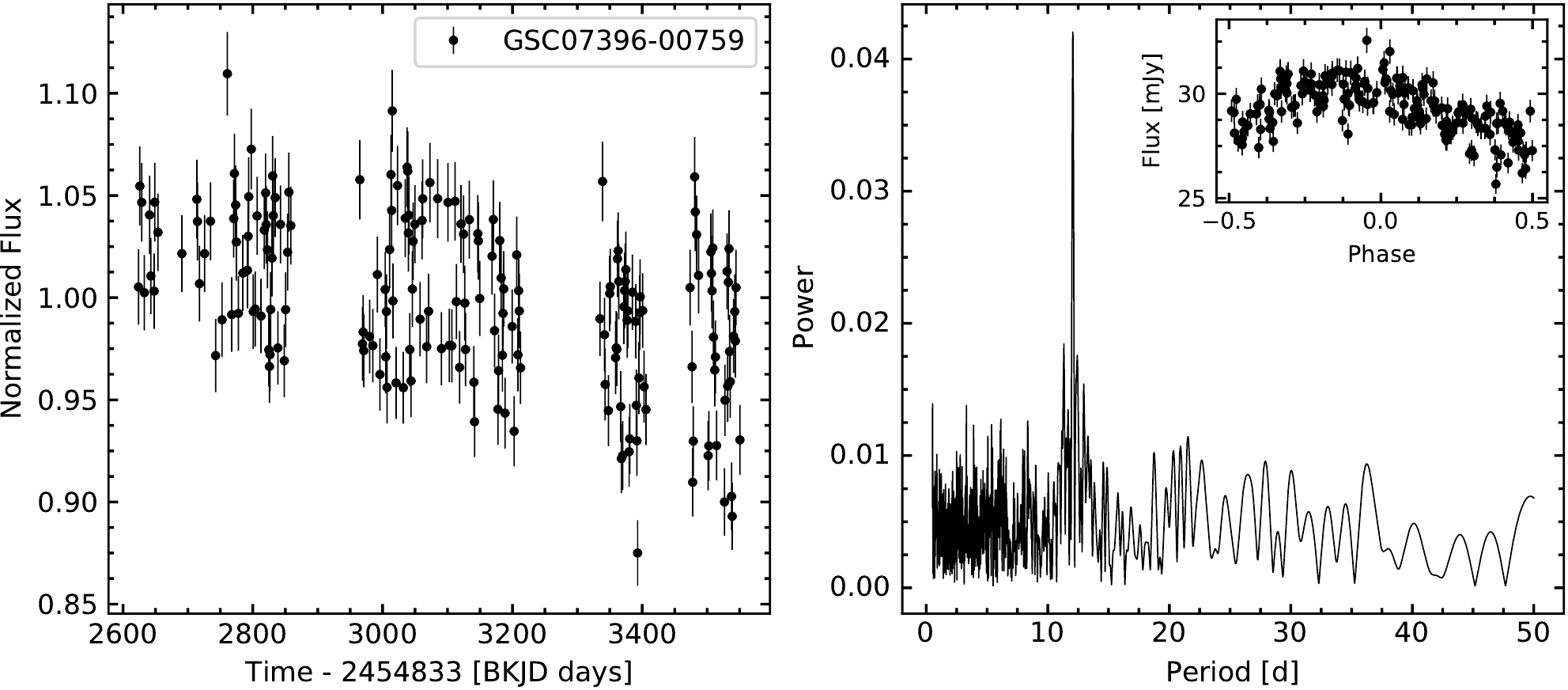}
\caption{Left panel: Normalized ASAS-SN light curve of \gsc, observed between 2016-03-10 and 2018-9-22. Right panel: Lomb-Scargle periodogram computed from these data, with an inset showing the light curve folded over the detected stellar rotation period $P_{\mathrm{rot}}=12.06$\,d.}
\label{fig:lc_asas-sn}
\end{figure*}

\clearpage
\section{Scattering angle}

As mentioned in Sec.~\ref{subsec:pp_ml}, we compute a grid of models to estimate whether we can reproduce the parameterized phase function, using the grain size and porosity as the only free parameters. However, as a consequence of the high inclination of the disk and the employed inner mask not all scattering angles between 0 and 180 degrees can be sampled. 
To account for this, we use the best-fit parameters determined for the disk, i.e. $r_{0}$, $\delta_{r}$, $i$, $\phi$, as well as the parameter of the employed masks to estimate range of valid scattering angle.
In Fig.~\ref{fig:scatter_ang} we show the results of this exercise which yielded observable scattering angle from 9.14\degree\ to 170.86\degree\ ($\theta_{\mathrm{min,max}}=90\degree\pm80.86\degree$). 

\begin{figure}[h] 
\centering
\includegraphics[width=9cm]{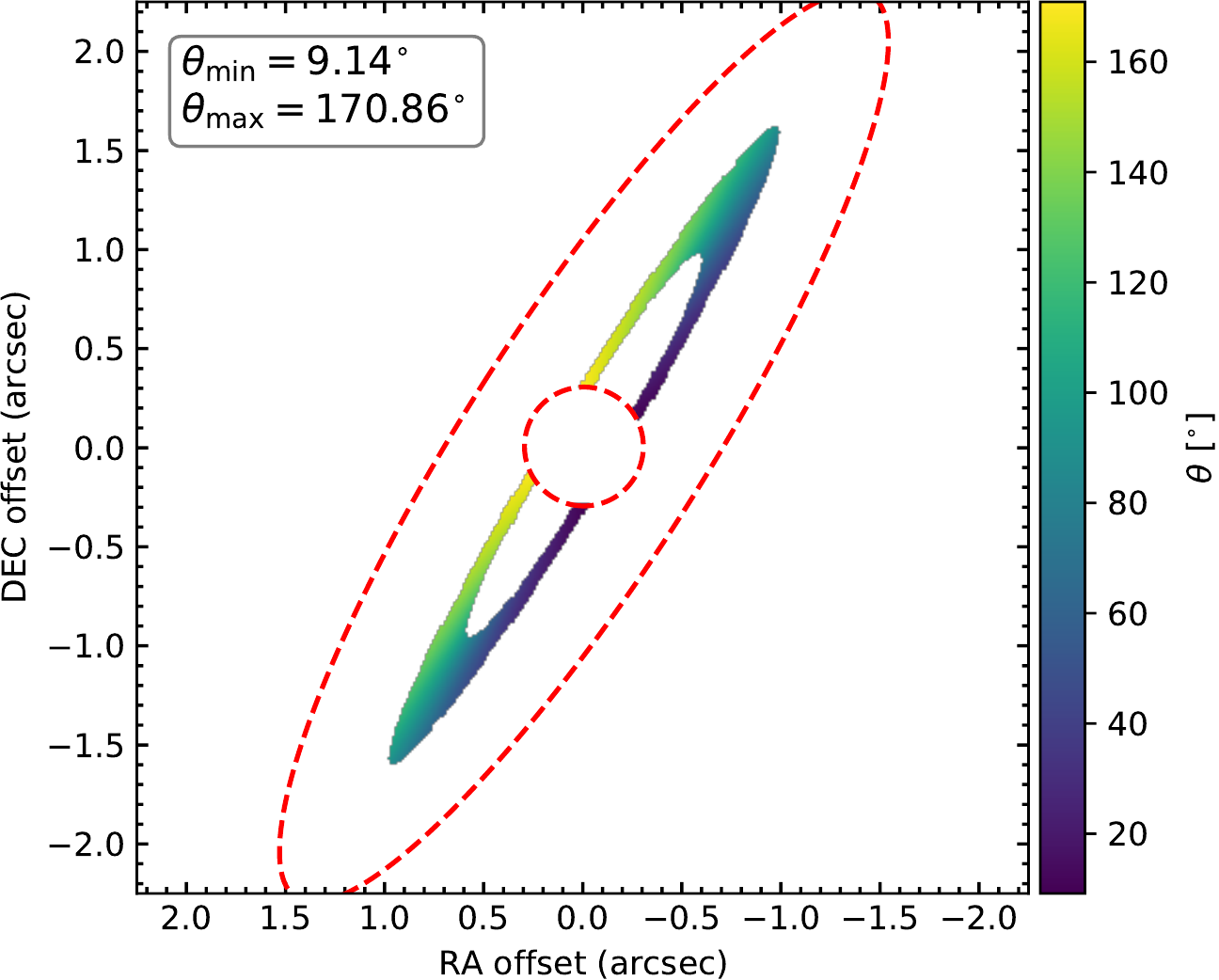}
\caption{Distribution of possible scattering angle as function of the model geometry obtained from our best-fit model of the disk. The inner and outer model boundaries are indicated by the circle and the ellipse in the image. Values within the circular mask or outside the ellipse are not considered in the modeling. North is to the top and east to the left.}
\label{fig:scatter_ang}
\end{figure}
\section{Model convergence}

The convergence and stability of our MCMC solution was assessed by estimating the maximum autocorrelation length among all parameters, i.e. the average autocorrelation time $\hat{\tau}$, after each iteration. 
The fitting was considered converged when the number of iterations is larger than 100 times the average autocorrelation time and its changes, between subsequent iterations, are less than 1\%. 
The evolution of the autocorrelation time as a function of the iteration step over the course of the modeling is presented in Fig.~\ref{fig:t_autocorr}.
At the end of the modeling, the average autocorrelation time was 79 steps and the mean acceptance fraction \citep{1992StaSc...7..457G} for our best-fitting model was 0.47.

\begin{figure}[h] 
\centering
\includegraphics[width=9cm]{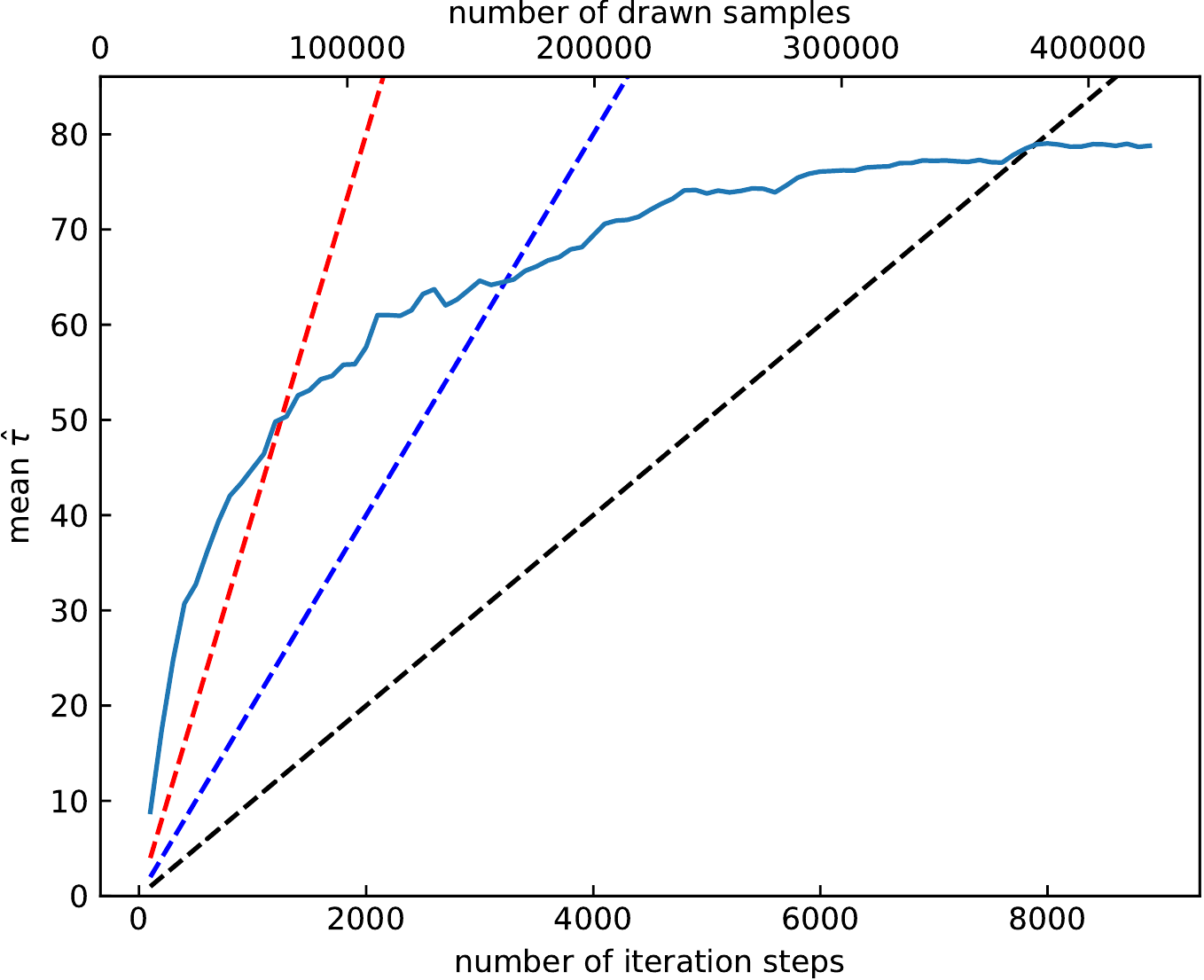}
\caption{Auto-correlation time as function of MCMC iteration steps. The estimated average auto-correlation time $\hat{\tau}$ is shown in blue (solid line). Also shown are the $\hat{\tau}=N_{\mathrm{iter}}/20$ (red), $N_{\mathrm{iter}}/50$ (blue), and $N_{\mathrm{iter}}/100$ (black) lines to indicate different possible levels of acceptance for the model parameter estimates. For this work, the model is considered converged if $\hat{\tau}< N_{\mathrm{iter}}/100$ and the change in consecutive estimated auto-correlation times $\tau$ is less than 1\%.}
\label{fig:t_autocorr}
\end{figure}

\section{Model comparison}
To explore the possibility that our radius for the disk around \gsc\ is overestimated, we re-run our grid of models with the reference radius fixed to $r_{0}=69.9$\,au, obtained by \cite{2018A&A...613L...6S} using forward modeling, to test whether we can find a solution that fits our observations.
The resulting best-fit model in this scenario is shown in Fig.~\ref{fig:data_model2}. In particular, we found a standard deviation of the radial distribution of $\delta_{r}\approx15.3\pm 0.2$\,au, an inclination of $i\approx 78.5\,\degree\pm0.3$ and a position angle $\phi\approx149.2\,\degree\pm0.1$.
For the analytical polarized phase function we found a coefficient $g=0.33$. 

\begin{figure*}[h] 
	\centering
	\includegraphics[width=18cm]{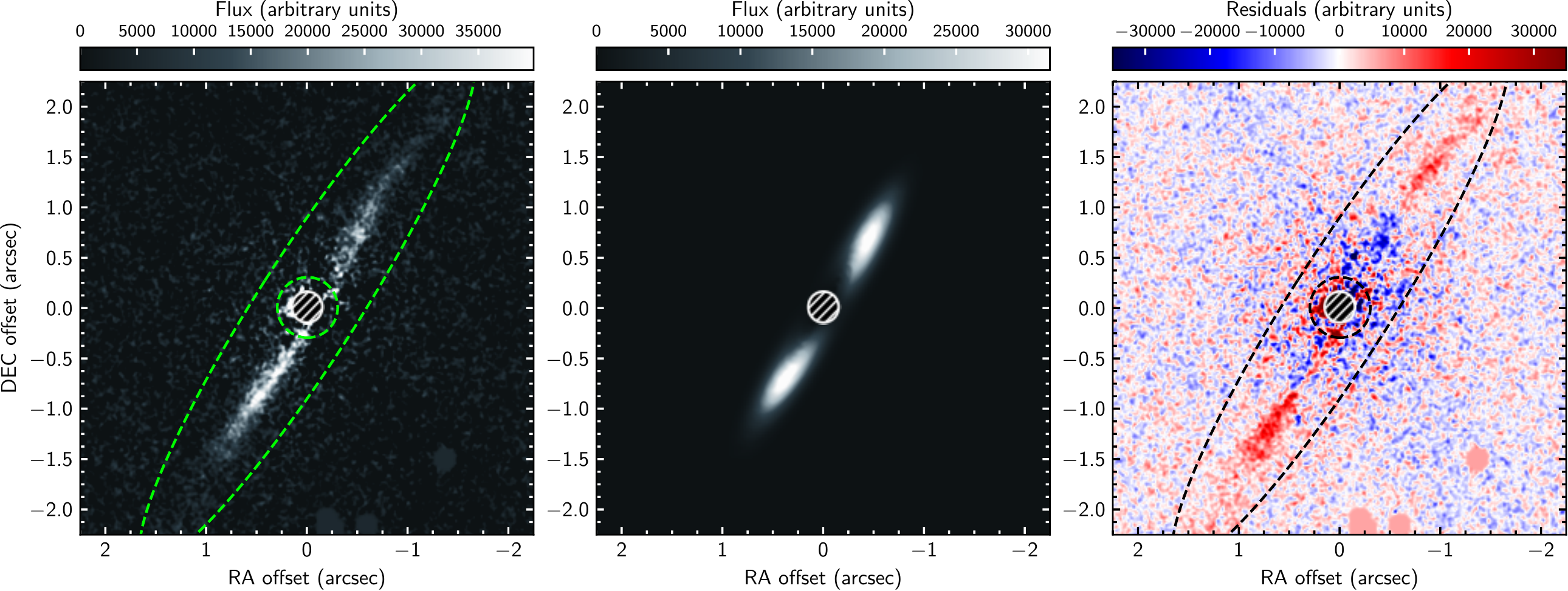}
	\caption{Observation and model images for \gsc. Shown from left to right are the measured $Q_{\phi}$, best-fit model and residual image obtained from our disk model using a linear scaling for each panel. The regions outside of the ellipse and within the circle (green dashed lines) are excluded from the $\chi^{2}$ calculation. The coronagraphic mask is indicated by the (shaded) circular region in each panel. North is to the top and east to left in each panel.}
	\label{fig:data_model2}
\end{figure*}

\section{Radius-luminosity relation}

We estimated the dependency of the disk radii on the stellar luminosity following the description in \citet{2018ApJ...859...72M}. In particular we used a power-law model where the belt locations $R_{i}$ (in au) are linked to their host star luminosities $L_{\star,i}$ (in \ls) through the form $R_{i}=R_{1\mathrm{L}_{\odot}}L_{\star,i}^{\alpha}+\epsilon_{i}$. The parameter $\epsilon_{i}$ represents the intrinsic scatter of the distribution, and is assumed to follow a Gaussian distribution with standard deviation $\sigma_{\mathrm{intr}} = f_{\Delta R} R_{i}$. 
The model was then applied to our sample which contains a total of 46 stars, including \gsc, with spectral types ranging from B9 to M3, ages between 10\,Myr and 2\,Gyr, and stellar luminosities reaching from 0.1\,\ls\ to 25\,\ls, presented in Table~\ref{table:4}.
The best solution to the power-law model, illustrated in Fig.~\ref{fig:r-l}, is determined using an affine invariant ensemble sample Monte-Carlo Markov Chain \citep[$\mathtt{emcee}$ package,][]{2013PASP..125..306F}. 
The 1-$\sigma$ confidence interval was estimated from a randomly drawn sample of the corresponding posterior probability distributions shown in Fig.~\ref{fig:corner_r-l}.
The 1D histograms presented in this figure represent the probability distributions of each parameter marginalized over the other two, while the contour maps represent the central 68.3\%, 95.5\%, and 99.73\% of the 2D probability distributions of different pairs of parameters, marginalized over the third.

\begin{table*}[h] 
	\caption{Properties of the sample of stars with disks resolved in scattered light.} 
	\label{table:4}	
	\centering	
	\begin{tabular}{lcccccccccccc}
		\hline\hline\\[-1em]
		Name & Sp. Type &      \lstar  &      Tel. &   Inst. & $\alpha_{\mathrm{in}}$ & $\alpha_{\mathrm{out}}$ & $r_{\mathrm{min}}$ & $r_{\mathrm{max}}$ & $\Delta r$ & $r_{0}$ &   FWHM & Ref. \\
		     &           &   (\ls)   &           &   &  &  & (au) & (au) & (au) & (au) & (au) &  \\
		\hline\\[-1em] 
		         *49Cet &      A1V &  15.71 &       VLT &  SPHERE &               2.6 &               -2.1 &                    &                    &            &   146.3 &  123.9 &    \hyperlink{cite.2017ApJ...834L..12C}{1} \\
		*alfPsA &      A3V &  16.50 &       HST &    STIS &                   &                    &              133.0 &              158.0 &            &   145.5 &   25.0 &    \hyperlink{cite.2013ApJ...775...56K}{2} \\
		*betPic &      A6V &   8.97 &    Gemini &     GPI &                   &                    &              23.61 &             138.84 &            &    81.2 &  115.2 &    \hyperlink{cite.2015ApJ...811...18M}{3} \\
		*gLup &      F5V &   3.33 &       HST &     ACS &               3.0 &               -2.0 &                    &                    &            &   110.3 &   91.9 &    \hyperlink{cite.2006ApJ...637L..57K}{4} \\
		GSC07396-00759 &      M2V &   0.14 &       VLT &  SPHERE &               2.8 &               -2.6 &                    &                    &            &    75.7 &   55.6 &    \hyperlink{cite.2018A\&A...613L...6S}{5} \\
		GSC07396-00759 &      M2V &   0.13 &       VLT &  SPHERE &                   &                    &                    &                    &       27.1 &   107.2 &   63.8 &  -- \\
		HD 104860 &       F8 &   1.18 &       HST &  NICMOS &               4.5 &               -3.9 &                    &                    &            &   118.4 &   57.8 &    \hyperlink{cite.2018ApJ...854...53C}{6} \\
		HD 106906 &      F5V &   7.14 &       VLT &  SPHERE &              10.0 &               -4.0 &                    &                    &            &    67.6 &   24.1 &    \hyperlink{cite.2016A\&A...586L...8L}{7} \\
		HD 107146 &      G2V &   0.98 &       HST &     ACS &               1.6 &               -2.8 &                    &                    &            &   135.5 &  117.0 &    \hyperlink{cite.2004ApJ...617L.147A}{8} \\
		HD 109573 &      A0V &  24.70 &  Magellan &   MagAO &              19.6 &               -6.0 &                    &                    &            &    81.4 &   17.5 &    \hyperlink{cite.2015ApJ...798...96R}{9} \\
		HD 110058 &      A0V &   9.36 &       VLT &  SPHERE &                   &                    &               20.0 &               65.0 &            &    42.5 &   45.0 &   \hyperlink{cite.2015ApJ...812L..33K}{10} \\
		HD 111161 &    A3III &   9.85 &    Gemini &     GPI &               2.5 &               -3.0 &                    &                    &            &    72.4 &   51.3 &   \hyperlink{cite.2020AJ....160...24E}{11} \\
		HD 111520 &      F5V &   2.75 &    Gemini &     GPI &                   &                    &               30.0 &              100.0 &            &    65.0 &   70.0 &   \hyperlink{cite.2016ApJ...826..147D}{12} \\
		HD 114082 &      F3V &   3.84 &       VLT &  SPHERE &               0.0 &               -4.0 &                    &                    &            &    29.6 &   34.0 &   \hyperlink{cite.2016A\&A...596L...4W}{13} \\
		HD 115600 &    F2.5V &   5.47 &    Gemini &     GPI &               7.5 &               -7.5 &                    &                    &            &    48.4 &   13.2 &   \hyperlink{cite.2015ApJ...807L...7C}{14} \\
		HD 117214 &      F6V &   6.13 &    Gemini &     GPI &               4.5 &               -4.5 &                    &                    &            &    60.2 &   27.1 &   \hyperlink{cite.2020AJ....160...24E}{10} \\
		HD 120326 &       F0 &   4.83 &       VLT &  SPHERE &              10.0 &               -5.0 &                    &                    &            &    60.6 &   18.6 &   \hyperlink{cite.2017A\&A...597L...7B}{15} \\
		HD 129590 &      G1V &   3.11 &       VLT &  SPHERE &               3.3 &               -2.4 &                    &                    &            &    73.1 &   53.1 &   \hyperlink{cite.2017ApJ...843L..12M}{16} \\
		HD 131835 &     A2IV &  10.94 &       VLT &  SPHERE &               8.2 &               -2.3 &                    &                    &            &    98.9 &   55.9 &   \hyperlink{cite.2017A\&A...601A...7F}{17} \\
		HD 143675 &      A5V &  11.44 &    Gemini &     GPI &               2.5 &               -3.0 &                    &                    &            &    46.3 &   32.8 &   \hyperlink{cite.2020AJ....160...24E}{10} \\
		HD 145560 &      F5V &   4.05 &    Gemini &     GPI &               3.5 &               -3.0 &                    &                    &            &    86.3 &   53.9 &   \hyperlink{cite.2020AJ....160...24E}{10} \\
		HD 146897 &    F2.5V &   3.38 &       VLT &  SPHERE &               5.0 &               -2.5 &                    &                    &            &    80.1 &   49.4 &   \hyperlink{cite.2017A\&A...607A..90E}{18} \\
		HD 15115 &     F4IV &   3.73 &       VLT &  SPHERE &               2.5 &               -4.5 &                    &                    &            &    94.4 &   55.0 &   \hyperlink{cite.2019A\&A...622A.192E}{19} \\
		HD 156623 &      A0V &  15.84 &    Gemini &     GPI &               1.5 &               -3.5 &                    &                    &            &    64.4 &   51.8 &   \hyperlink{cite.2020AJ....160...24E}{10} \\
		HD 15745 &      F2V &   4.21 &       HST &     ACS &                   &                    &              128.0 &              450.0 &            &   289.0 &  322.0 &   \hyperlink{cite.2007ApJ...671L.161K}{20} \\
		HD 157587 &      F5V &   4.38 &    Gemini &     GPI &                   &                    &               80.0 &              213.0 &            &   146.5 &  133.0 &   \hyperlink{cite.2016AJ....152..128M}{21} \\
		HD 160305 &      F9V &   1.69 &    SPHERE &   IRDIS &              10.1 &               -7.1 &                    &                    &            &    92.4 &   22.8 &   \hyperlink{cite.2019A\&A...626A..95P}{22} \\
		HD 172555 &      A7V &   7.72 &       VLT &  SPHERE &               2.3 &               -9.8 &                    &                    &            &    10.9 &    4.8 &   \hyperlink{cite.2018A\&A...618A.151E}{23} \\
		HD 181327 &    F5.5V &   2.88 &       HST &  NICMOS &                   &                    &                    &                    &       36.0 &    86.3 &   36.0 &   \hyperlink{cite.2006ApJ...650..414S}{24} \\
		HD 191089 &      F5V &   2.74 &    Gemini &     GPI &               4.9 &               -6.1 &                    &                    &            &    43.9 &   16.2 &   \hyperlink{cite.2019ApJ...882...64R}{25} \\
		HD 192758 &      F0V &   5.43 &       HST &  NICMOS &               2.9 &               -2.0 &                    &                    &            &   109.2 &   91.9 &    \hyperlink{cite.2018ApJ...854...53C}{5} \\
		HD 202628 &      G5V &   0.98 &       HST &    STIS &                   &                    &                    &                    &       60.0 &   175.3 &   60.0 &   \hyperlink{cite.2016AJ....152...64S}{26} \\
		HD 202917 &      G7V &   0.67 &       HST &    STIS &                   &                    &                    &                    &       13.2 &    62.8 &   13.2 &   \hyperlink{cite.2016AJ....152...64S}{25} \\
		HD 207129 &      G2V &   1.21 &       HST &    STIS &                   &                    &                    &                    &       72.3 &   149.4 &   72.3 &   \hyperlink{cite.2016AJ....152...64S}{25} \\
		HD 30447 &      F3V &   3.73 &       HST &  NICMOS &                   &                    &               60.0 &              200.0 &            &   130.0 &  140.0 &   \hyperlink{cite.2014ApJ...786L..23S}{27} \\
		HD 32297 &       A5 &   8.47 &    SPHERE &   IRDIS &              10.0 &               -4.0 &                    &                    &            &   138.6 &   49.3 &   \hyperlink{cite.2019A\&A...630A..85B}{28} \\
		HD 35650 &      K6V &   0.13 &       HST &  NICMOS &               5.0 &               -5.0 &                    &                    &            &    55.1 &   22.4 &   \hyperlink{cite.2016ApJ...817L...2C}{29} \\
		HD 35841 &      F5V &   2.43 &    Gemini &     GPI &               3.8 &               -3.0 &                    &                    &            &    56.5 &   34.3 &   \hyperlink{cite.2018AJ....156...47E}{30} \\
		HD 36546 &       B9 &  15.86 &    Subaru &  HiCIAO &               3.0 &               -3.0 &                    &                    &            &    90.1 &   59.6 &   \hyperlink{cite.2017ApJ...836L..15C}{31} \\
		HD 377 &      G2V &   1.16 &       HST &  NICMOS &               5.0 &               -5.0 &                    &                    &            &    87.8 &   35.7 &   \hyperlink{cite.2016ApJ...817L...2C}{28} \\
		HD 53143 &      K1V &   0.59 &       HST &     ACS &                   &                    &               55.0 &              110.0 &            &    82.5 &   55.0 &    \hyperlink{cite.2006ApJ...637L..57K}{3} \\
		HD 61005 &      G8V &   0.64 &       VLT &  SPHERE &               5.0 &               -2.7 &                    &                    &            &    65.6 &   38.5 &   \hyperlink{cite.2016A\&A...591A.108O}{32} \\
		V*AU Mic &      M1V &   0.09 &       HST &     ACS &                   &                    &                7.5 &              150.0 &            &    78.8 &  142.5 &   \hyperlink{cite.2005AJ....129.1008K}{33} \\
		V*CE Ant &     M3.2 &   0.11 &       VLT &  SPHERE &               5.0 &               -1.5 &                    &                    &            &    30.3 &   26.7 &   \hyperlink{cite.2018A\&A...617A.109O}{34} \\
		V*NZ Lup &       G2 &   2.07 &    SPHERE &   IRDIS &               7.0 &               -5.0 &                    &                    &            &    95.2 &   33.7 &   \hyperlink{cite.2019A\&A...625A..21B}{35} \\
		V*V1249 Cen &    M0.5V &   0.23 &       HST &  NICMOS &               5.0 &               -5.0 &                    &                    &            &    79.6 &   32.4 &   \hyperlink{cite.2016ApJ...817L...2C}{28} \\
		V*V419 Hya &      K1V &   0.38 &       HST &     ACS &                   &                    &               43.0 &              110.0 &            &    76.5 &   67.0 &   \hyperlink{cite.2011AJ....142...30G}{36} \\
		
		\hline 
	\end{tabular}
	
	\tablefoot{$\alpha_{\mathrm{in}}$: Inner radial index. $\alpha_{\mathrm{out}}$: Outer radial index. $r_{\mathrm{min}}$: Inner disk radius detected from scattered light. $r_{\mathrm{max}}$: Outer disk radius detected from scattered light. $\Delta r$: Width of the radial distribution. $r_{0}$: Reference radius and center of the radial distribution. FWHM: Full width at half maximum.}
	
	\tablebib{ (--) This work; (1) \citet{2017ApJ...834L..12C}, (2) \citet{2013ApJ...775...56K}, (3) \citet{2015ApJ...811...18M}, (4) \citet{2006ApJ...637L..57K}, (5) \citet{2018A&A...613L...6S}, (6) \citet{2018ApJ...854...53C}, (7) \citet{2016A&A...586L...8L}, (8) \citet{2004ApJ...617L.147A}, (9) \citet{2015ApJ...798...96R}, (10) \citet{2015ApJ...812L..33K}, (11) \citet{2020AJ....160...24E}, (12) \citet{2016ApJ...826..147D}, (13) \citet{2016A&A...596L...4W}, (14) \citet{2015ApJ...807L...7C}, (15) \citet{2017A&A...597L...7B}, (16) \citet{2017ApJ...843L..12M}, (17) \citet{2017A&A...601A...7F}, (18) \citet{2017A&A...607A..90E}, (19) \citet{2019A&A...622A.192E}, (20) \citet{2007ApJ...671L.161K}, (21) \citet{2016AJ....152..128M}, (22) \citet{2019A&A...626A..95P}, (23) \citet{2018A&A...618A.151E}, (24) \citet{2006ApJ...650..414S}, (25) \citet{2019ApJ...882...64R}, (26) \citet{2016AJ....152...64S}, (27) \citet{2014ApJ...786L..23S}, (28) \citet{2019A&A...630A..85B}, (29) \citet{2016ApJ...817L...2C}, (30) \citet{2018AJ....156...47E}, (31) \citet{2017ApJ...836L..15C}, (32) \citet{2016A&A...591A.108O}, (33) \citet{2005AJ....129.1008K}, (34) \citet{2018A&A...617A.109O}, (35) \citet{2019A&A...625A..21B}, (36) \citet{2011AJ....142...30G}
	}
\end{table*}
\begin{figure}[h] 
	\centering
	\includegraphics[width=9cm]{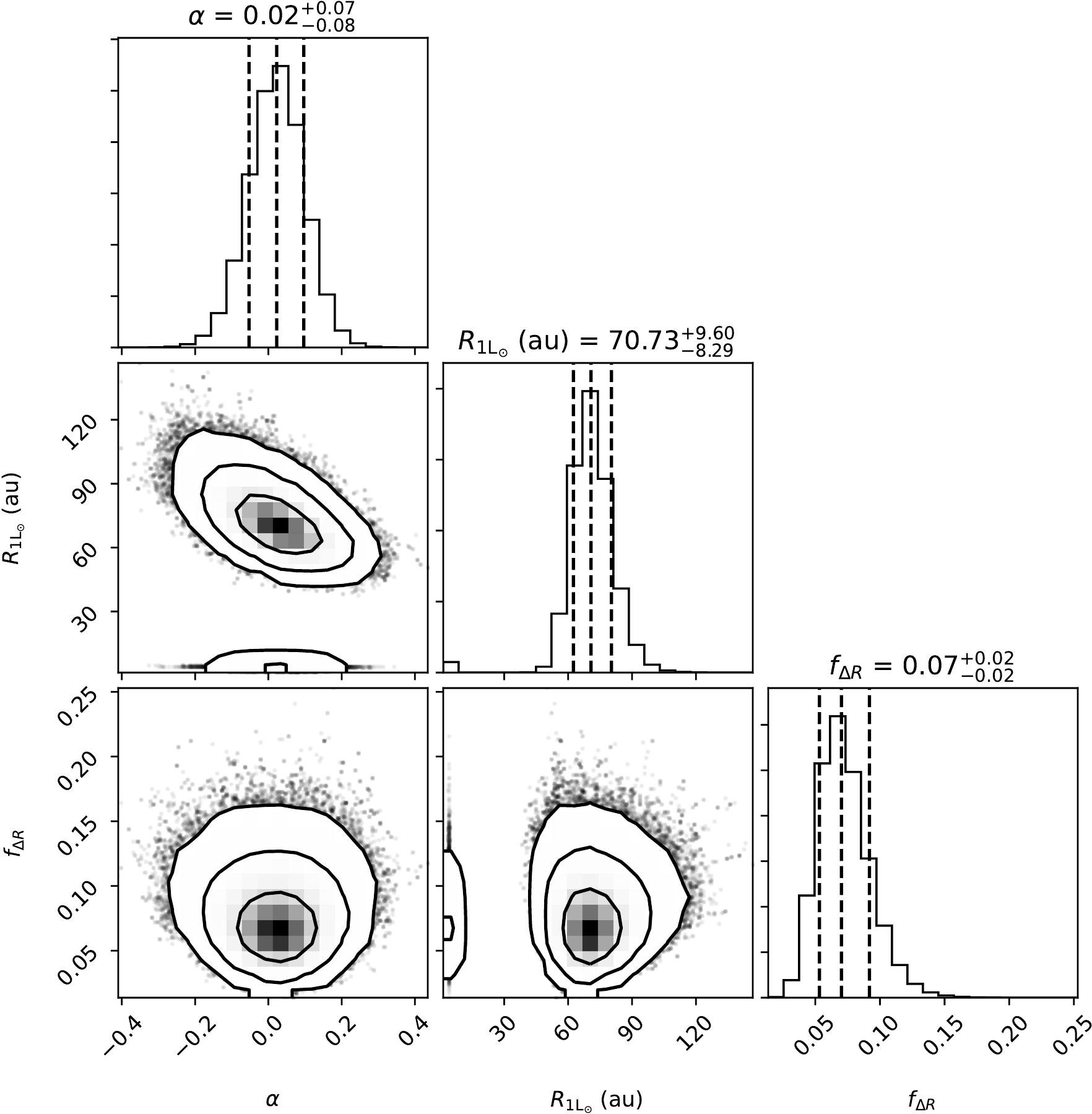}
	\caption{Projected posterior distribution and density plots of the free parameter (slope $\alpha$, intercept $R_{1\mathrm{L}_{\odot}}$, and fractional intrinsic scatter $f_{\Delta R}$) of the power-law fitted to the sample data points. The plot additionally shows the 50\,\%, 16\,\%, and 84\,\% quartiles (vertical dashed lines), representing the distributions median and the 1\,$\sigma$ uncertainties (lower and upper bound), respectively. The 1D histograms represent the probability distributions of each parameter marginalized over the other two. The contour maps represent the central 68.3\%, 95.5\%, and 99.73\% of the 2D probability distributions of different pairs of parameters, marginalized over the third.}
	\label{fig:corner_r-l}
\end{figure}

\end{appendix}
\end{document}